\documentclass[aps,prb,reprint,amsmath,amssymb,superscriptaddress,showpacs]{revtex4-1}
\usepackage{txfonts}
\usepackage{bm}
\usepackage{graphicx}
\usepackage{color}
\usepackage{dcolumn}
\usepackage{kotex}
\usepackage{float}
\usepackage{hyperref}
\usepackage{amsmath}
\usepackage{siunitx}
\usepackage{multirow}
\newcommand{\rom}[1]{\uppercase\expandafter{\romannumeral #1\relax}}

\begin{document}

\title{Hybridization and Decay of Magnetic Excitations in Two-Dimensional Triangular Lattice Antiferromagnets}

\author{Taehun Kim}
\author{Kisoo Park}
\author{Jonathan C. Leiner}
\author{Je-Geun Park}\email{jgpark10@snu.ac.kr}
\affiliation{Center for Correlated Electron Systems, Institute for Basic Science (IBS), Seoul 08826, Republic of Korea}
\affiliation{Department of Physics and Astronomy, Seoul National University, Seoul 08826, Republic of Korea}

\begin{abstract}
	Elementary quasiparticles in solids such as phonons and magnons occasionally have nontrivial interactions between them, as well as among themselves. As a result, their energy eigenvalues are renormalized, the quasiparticles spontaneously decay into a multi-particle continuum state, or they are hybridized with each other when their energies are close. As discussed in this review, such anomalous features can appear dominantly in quantum magnets but are not, a priori, negligible for magnetic systems with larger spin values and noncollinear magnetic structures. We review the unconventional magnetic excitations in two-dimensional triangular lattice antiferromagnets and discuss their implications on related issues.
\end{abstract}

\maketitle

\section{Introduction}
\label{sec:Intro}
Magnons and phonons are the elementary quasiparticles that arise from the respective underlying magnetic and crystalline orders in materials. With energy scales ranging from $\sim$1-100 meV, both magnons and phonons can be well described within the scheme of linearized theories such as linear spin wave theory (LSWT) and density functional theory (DFT) calculations, respectively. These are standard models in modern condensed matter physics, and they both assume that there is very little, if any, coupling among these two fundamental excitations.

Very interesting and non-trivial behavior arise when these two quasiparticles interact with each other. Spin-lattice coupling\cite{Proc.R.Soc.Lond.A.264.458.1961} is the main mechanism, which facilitates the interaction between magnons and phonons.\cite{NatCommun.7.13146,PhysRev.110.836,J.PhysC.5.2769}  One form of the spin-lattice coupling is so-called exchange-striction, the result of modulations in exchange interactions due to atomic displacements.\cite{PhysRevLett.105.037205} Under the right conditions, such as noncollinear spin order, the exchange-striction mechanism can produce significant changes in the ground states and/or excitations that are otherwise forbidden in collinear magnets.\cite{PhysRevLett.100.077201} With a sufficiently large spin-lattice coupling strength,\cite{PhysRevB.71.060407,PhysRevLett.103.067204} magnon-phonon hybridization can significantly influence both the magnon and phonon spectra measured by experiments below the magnetic ordering temperature. The resulting magneto-phonon mode born of this hybridization can produce additional peaks in the magnetic excitation spectra,\cite{NatCommun.7.13146} which recently has been successfully measured by experiments and modeled by properly considering the in-plane Mn-O bond length changes from exchange-striction for several hexagonal manganites.\cite{NatCommun.7.13146} 

\begin{figure}
	\includegraphics[width=1\columnwidth,clip]{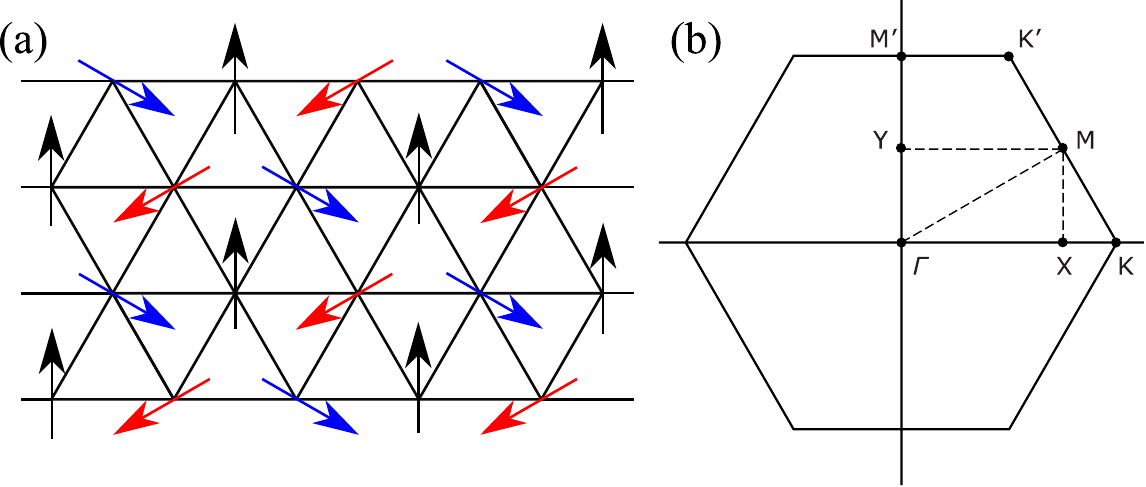}
	\caption{(a) 120$^\circ$ spin ordered state in triangular lattice. (b) 1\textsuperscript{st} Brillouin zone in triangular lattice and labels for high symmetric points.}
	\label{f1}
\end{figure}

The breakdown or decay of the quasiparticles due to their interactions with a multi-particle continuum can be more easily realized in strongly correlated electron systems.\cite{Nature.440.187,NatPhys.12.224} For instance, noncollinear magnetic structures allow for three-magnon interactions, leading to the spontaneous decay of magnons when the kinematic conditions are satisfied.\cite{PhysRevB.79.144416,RevModPhys.85.219} These intrinsic zero-temperature decays are significantly enhanced (i.e. there are singularities in the decays) at certain momentum transfers, as found by calculations taking into account the anharmonicity of the spin waves.\cite{PhysRevB.79.144416,RevModPhys.85.219,PhysRevLett.97.207202,PhysRevB.88.094407,PhysRevB.93.099901} Essentially, the noncollinear magnetic structure breaking the O(3) symmetry allows a three-magnon interaction term in the Hamiltonian, and decay channels are made possible through which single-magnons spontaneously decay into two-magnon states. These decay processes result in the enhanced linewidth of the magnon modes observed by experiments.\cite{PhysRevLett.111.257202}

Inelastic neutron scattering (INS) is often an ideal probe for investigating the collective behavior of both magnons and phonons.\cite{npj.2.63,NatPhys.1,PhysRevMaterials.2.024404,JKPS.63.333} It allows for the direct comparison between experimental data and theoretical calculations of the dynamic spin-spin correlation function (or dynamical structure factor). On the other hand, inelastic X-ray scattering (IXS) can provide complementary information on the magnon-phonon coupling, as it is exclusively sensitive to the phonon excitations.\cite{J.Phys.Chem.Solid.61.461} By using both inelastic neutron and X-ray scattering techniques, the full spin Hamiltonian, including exchange-striction type linear magnon-phonon coupling and their higher order interaction terms, has been unambiguously determined in several key cases, which we will highlight in this review. 

For two-dimensional (2D) triangular lattice antiferromagnet (TLAF) compounds\cite{Can.J.Phys.75.605} with noncollinear magnetic ground states, these magnon-phonon couplings and magnon self-interactions can become non-negligible and sometimes quite strong. Extensive studies on hexagonal RMnO${}_{3}$ (rare-earth atom R = Y, Lu or Ho) compounds\cite{PhysRevB.76.144406,PhysRevB.82.184420,PhysRevLett.111.257202,PhysRevLett.94.087601,PhysRevB.97.201113} have demonstrated beyond any doubt the existence of new hybrid excitations arising from these couplings, as well as one of the largest known magnon-phonon coupling strengths.\cite{NatCommun.7.13146} Several other variations of triangular lattice compounds such as the spatially anisotropic TLAF Cs$_2$CuCl$_4$\cite{PhysRevB.64.094425,PhysRevLett.79.151,PhysRevB.73.184403,NatPhys.3.790} and the spin-1/2 equilateral TLAF Ba$_3$CoSb$_2$O$_9$\cite{PhysRevB.91.134423,NatCommun.9.2666,PhysRevLett.116.087201,PhysRevLett.110.267201} allow for the effects of systematic perturbations on the spin-lattice coupling to be discerned. We will also discuss other examples of systems exhibiting magneto-phonon modes such as two delafossites CuCrO$_2$\cite{PhysRevB.94.104421,JPCM.24.036003,PhysRevB.84.094448,JPSJ.84.074708,JPSJ.78.113710,PhysRevB.81.104411,JKPS.62.2168} and LiCrO$_2$\cite{NatCommun.7.13547,JPCM.7.6869,PhysRevB.79.184411}.

With regard to potential applications, magnon-phonon coupling is expected to play an important role in several diverse fields, in particular spintronics.\cite{PhysRevB.97.180301} Typically, phonons have a large group velocity, and so they travel relatively swiftly through materials. However, when magnons are required for the spintronic devices, the fast-traveling phonons could be converted into magnons ``on-demand'' provided their coupling behavior is sufficiently understood.\cite{PhysRevB.96.100406,PhysRevB.92.214437} Recent results with yttrium iron garnet (YIG) film have demonstrated that magneto-elastic waves are capable of driving magnetic bubble domains (i.e. curved domain walls), and therefore harnessing a versatile form of spin-momentum for use in new spintronic architectures.\cite{Proc.NAS.USA.112.8977}

In this review article, the focus will be on conveying the essential groundwork describing both the hybridization and decay of magnon excitations (See the Table \ref{t1} for the summary of theoretical models). Particular attention will be devoted to frustrated magnets, where competing ground states can lead to greater sensitivity to the effects from magnons and phonons coupling to each other.\cite{PhysRevB.68.134424,PhysRevB.71.212406,PhysRevB.84.024414,PhysRevLett.96.057201} This comprehensive overview of the present knowledge of these phenomena in 2D TLAF systems should thoroughly motivate path forward in addressing the remaining challenges of this arena, such as understanding the impact magnon-phonon coupling can have on three-dimensional lattice compounds with noncollinear magnetic structures.\cite{PhysRevLett.118.117201,PhysRevLett.119.057203}

\begin{figure*}
	\includegraphics[width=1\textwidth,clip]{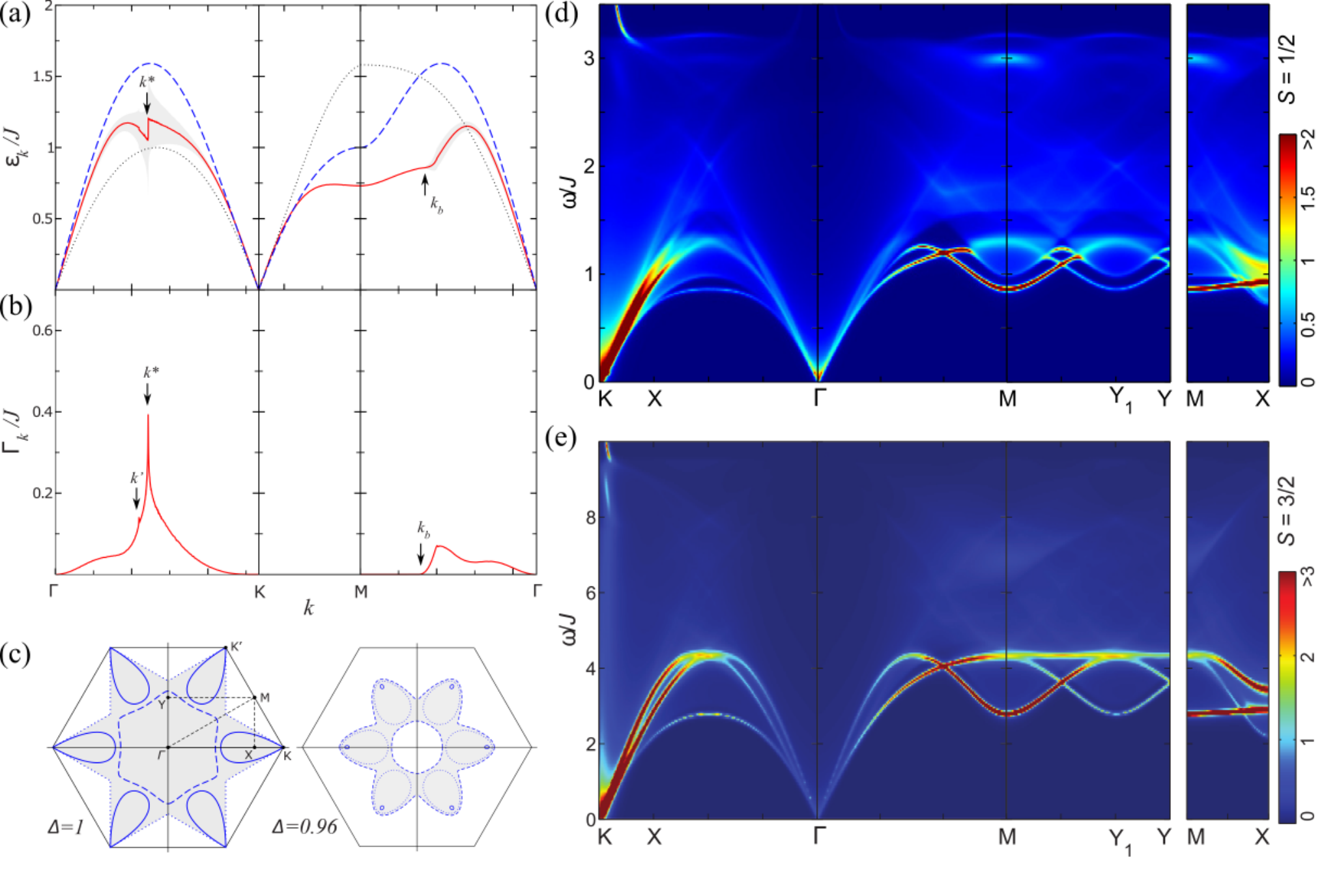}
	\caption{(a) Calculated spin wave dispersion and (b) decay rate of a triangular lattice with $S$ =  1/2. Blue dashed and red solid lines are calculated from LSWT approximation and with $1/S$ order corrections. Dotted lines represent the minimum of the two-magnon continuum calculated by LSWT. Gray area represents the width of the spectral peaks due to damping. (c) Sketch of the decay region allowed in the reciprocal space of 1\textsuperscript{st} Brillouin zone for two cases: one is isotropic exchange with $\Delta=1$ and the other anisotropic exchange with $\Delta=0.96$. (d),(e) Intensity plots of the momentum and energy dependence of the total dynamical strucure factor of 2D TLAF with $S$ =  1/2 and 3/2. Reprinted with permission from Chernyshev et al.\cite{PhysRevB.79.144416} and Mourigal et al.\cite{PhysRevB.88.094407,PhysRevB.93.099901} (Copyright © American Physical Society).}
	\label{f2}
\end{figure*}

\section{Theoretical Overview}
\label{sec:2}
\subsection{Spin waves of triangular lattice antiferromagnet (TLAF)}
\label{sec:2.1}
\subsubsection{Magnetic ground state}
\label{sec:2.1.1}
As one of the most actively studied frustrated magnets, the 2D TLAF offers a complex phase diagram of novel phenomena.\cite{Can.J.Phys.75.605} Here we focus on the 120$^\circ$ magnetic order of the well-known noncollinear magnetic ground state. In the semiclassical (large S) triangular lattice, the 120$^\circ$ magnetic order is well stabilized by the combination of the competing exchange interactions and low-dimensionality (see Fig. \ref{f1}(a)). For the quantum spin-1/2 case, it was originally speculated that strong quantum fluctuations would eventually destroy the long-range order and lead to the now well-known, yet elusive,  resonating valence bond liquid state.\cite{MRB.8.153} However, a 120$^\circ$ magnetic order is still found to survive even in the quantum spin model, as is evident in several theoretical works\cite{PhysRevLett.96.057201,PhysRevLett.60.2531,PhysRevB.40.2727,PhysRevLett.69.2590,JPSJ.61.983,JPCM.6.8891,PhysRevLett.82.3899,PhysRevLett.99.127004} and numerous experimental realizations.\cite{PhysRevLett.116.087201,PhysRevLett.109.267206,PhysRevB.90.014403}

\subsubsection{Effect of noncollinearity}
\label{sec:2.1.2}
The aforementioned noncollinearity gives rise to novel properties in the spin excitations, which are otherwise not allowed in conventional collinear magnets. For example, the SO(2) spin rotational symmetry of the collinear magnets would forbid possible mixing between the transverse ($S^{x,y}$) and longitudinal ($S^z$) fluctuations. The transverse (longitudinal) fluctuations carry an odd (even) number of magnons, and so the lowest-order magnon-magnon interaction terms in collinear magnets are of quartic order, making the higher order corrections of the spin waves intrinsically weak.\cite{PhysRevB.3.961,PhysRev.117.117,PhysRev.102.1217,JPCM.22.216003,PhysRevB.71.184440,PhysRevB.72.014403,PhysRevB.47.7961} On the other hand, the spin rotational symmetry is completely broken in noncollinear antiferromagnets, and thus one can expect odd numbers of magnon terms to be present in spin Hamiltonian, cubic anharmonic terms in particular. Such cubic anharmonicity would make significant differences in the magnon spectra obtained from the harmonic approximation, especially for a spin-1/2 case. In this regard, the spin-1/2 TLAF provides a useful platform for the study of the higher order effects in quantum magnets with a noncollinear magnetic order because of the strong correction terms, whose linear terms are vanished in TLAF and the cubic terms are purely survived. We provide the further detailed analysis of the role of cubic anharmonicity in the next section.

\subsubsection{Magnetic Hamiltonian}
\label{sec:2.1.3}
The basic Heisenberg Hamiltonian for a 2D triangular lattice is given by
\[
{\mathcal{H}}_{Heis}\mathrm{=}J\sum_{\left\langle ij\right\rangle }{{\boldsymbol{S}}_i\cdot {\boldsymbol{S}}_j}, \tag{1} \label{Heis}
\] 
where ${\boldsymbol{S}}_i$ is a spin operator at ${\boldsymbol{r}}_i$, and $J$ is an exchange interaction running over the nearest neighbor $\left\langle ij\right\rangle $ pairs. Here we use the same notations as in Chernyshev et al\cite{PhysRevB.79.144416}, in which the detailed explanation and derivation are given more explicitly. We consider all spins lying in the $xz$ plane and rewrite Eq. \eqref{Heis} using the transformation from a laboratory frame $\left(x_0,\ y_0,\ z_0\right)$ to the local frame $\left(x,\ y,\ z\right)$,
\begin{align*}
{\mathcal{H}}_{Heis}&\mathrm{=}J\sum_{\left\langle ij\right\rangle }S^y_iS^y_j+{\mathrm{cos} {\theta }_{ij}\ }\left(S^z_iS^z_j+S^x_iS^x_j\right)\\
	&+{\mathrm{sin} {\theta }_{ij}\ }\left(S^z_iS^x_j-S^x_iS^z_j\right), \tag{2} \label{localHeis}
\end{align*}
where ${\theta }_{ij}\mathrm{=}\boldsymbol{Q}\cdot \left({\boldsymbol{r}}_i-{\boldsymbol{r}}_j\right)=\pm 120\mathrm{{}^\circ }$ is a relative angle between ${\boldsymbol{S}}_i$ and ${\boldsymbol{S}}_j$, and $\boldsymbol{Q}=\left(4\pi/3\:0\right)$ is the ordering wave vector of the 120$\mathrm{{}^\circ }$ spiral order.

\subsubsection{$1/S$ expansion and harmonic approximation}
\label{sec:2.1.4}
Using the Holstein-Primakoff (HP) transformation,\cite{PhysRev.58.1098} the spin components in Eq. \eqref{localHeis} are bosonized:$\ S^z_i\mathrm{=}S\mathrm{-}a^{\dagger }_ia_i$ and $S^-_i=a^{\dagger }_i\sqrt{2S-a^{\dagger }_ia_i}$ $\mathrm{,\ where\ }S^{\pm }_i=S^x_i\pm iS^y_i$. Expansion in powers of ${1}/{S}$ produces the following bosonic Hamiltonian: ${\mathcal{H}}_{Heis}={\mathcal{H}}_0+{\mathcal{H}}_1+{\mathcal{H}}_2+{\mathcal{H}}_3+\mathrm{\cdots }$, where ${\mathcal{H}}_n$ denotes the Hamiltonian that contains the \textit{n}$^{th}$ powers of the bosonic operators. ${\mathcal{H}}_0$ describes the classical energy of this system. Since the spin waves represent deviations from the magnetic ground state, ${1}/{S}$ expansion requires the cancellation of ${\mathcal{H}}_1$.  In the LSWT, the quadratic form can be written as follows:
\[
{\mathcal{H}}_2=\sum_{k}\left[{A_ka_k^\dagger a_k-\frac{1}{2}B_k\left({a_k^{\dagger}a_{-k}^{\dagger}+a_{-k}a_k}\right)}\right], \tag{3} \label{H2}
\]
where $A_k=3JS\left(1+\frac{1}{2}\gamma_k\right)$, $B_k=\frac{9}{2}JS\gamma_k$, and ${\gamma }_{\boldsymbol{k}}\mathrm{=}\frac{1}{6}\sum_{\boldsymbol{\delta }}{e^{i\boldsymbol{k}\cdot \boldsymbol{\delta }}}$. It can then be diagonalized using the canonical Bogoliubov transformation, $a_{\boldsymbol{k}}\mathrm{=}u_{\boldsymbol{k}}{\alpha }_{\boldsymbol{k}}+v_{\boldsymbol{k}}{\alpha }^{\dagger }_{\boldsymbol{-}\boldsymbol{k}}$, resulting in the harmonic magnon Hamiltonian:
\[
{\mathcal{H}}_2=\sum_{\boldsymbol{k}}{{\varepsilon }_{\boldsymbol{k}}}{\alpha }^{\dagger }_{\boldsymbol{k}}{\alpha }_{\boldsymbol{k}}, \tag{4} \label{H2_diagonalized} 
\] 
where ${\varepsilon }_{\boldsymbol{k}}=3JS\sqrt{\left(1-{\gamma }_{\boldsymbol{k}}\right)\left(1+2{\gamma }_{\boldsymbol{k}}\right)}$ is the magnon energy in the harmonic approximation.

\subsubsection{Magnon-magnon interaction}
\label{sec:2.1.5}
In noncollinear TLAFs, the $1/S$ expansion of the magnon spectra has significant deviations from its harmonic approximation, which are ascribed to the presence of a sizable magnon-magnon interaction. For example, an early theoretical report by Chubukov et al. demonstrated that the first order correction in the $1/S$ expansion can be non-negligible for the $S$ = 1/2 case.\cite{JPCM.6.8891} Subsequently, Chernyshev et al.\cite{PhysRevB.79.144416} conducted a comprehensive study on the quantum corrections to the spin wave dispersion and the magnon damping due to their anomalous decay rate. Zheng et al. also demonstrated the effects of the higher-order term on the magnon dispersion curve using a quantum Monte-Carlo method.\cite{PhysRevLett.96.057201} Later, Mourigal et al. provided the explicit calculation of the dynamical structure factor using the $1/S$ formalism with help of a Dyson equation.\cite{PhysRevB.88.094407,PhysRevB.93.099901} For more detailed information, see the review written by Zhitomirsky et al.\cite{RevModPhys.85.219}

As shown in Sect. \ref{sec:2.1.4}, a cubic interaction term ${\mathcal{H}}_3$ is given by
\[
{\mathcal{H}}_3=J\sqrt{\frac{S}{2}}\sum_{\left\langle ij\right\rangle }{{\mathrm{sin} {\theta }_{ij}\ }\left(a^{\dagger }_ia_i\left(a^{\dagger }_j+a_j\right)-a^{\dagger }_ja_j\left(a^{\dagger }_i+a_i\right)\right)}. \tag{5} \label{H3_a}
\] 
With help of the Fourier and Bogoliubov transformations, one can obtain the general form of the cubic interaction term:
\begin{align*}
{\mathcal{H}}_3&\mathrm{=}\frac{1}{2!}\sum_{\boldsymbol{q},\boldsymbol{k}}{{\mathrm{\Gamma }}_1\left(\boldsymbol{q}\boldsymbol{;}\boldsymbol{k}\right)\left(\alpha^{\dagger }_{\boldsymbol{k}\boldsymbol{-}\boldsymbol{q}}\alpha^{\dagger }_{\boldsymbol{q}}\alpha_{\boldsymbol{k}}+\mathrm{h}.\mathrm{c}.\right)}\\
&+\frac{1}{3!}\sum_{\boldsymbol{q},\boldsymbol{k}}{{\mathrm{\Gamma }}_2\left(\boldsymbol{q},\boldsymbol{k}\right)\left(\alpha^{\dagger }_{\boldsymbol{k}\boldsymbol{-}\boldsymbol{q}}\alpha^{\dagger }_{\boldsymbol{q}}\alpha^{\dagger }_{\boldsymbol{k}}+\mathrm{h}\mathrm{.c.}\right)}, \tag{6} \label{H3_alpha}
\end{align*} 
where ${\mathrm{\Gamma }}_1\left(\boldsymbol{q}\boldsymbol{;}\boldsymbol{k}\right)$ is the vertex function that describes the decay of one-magnon into the two-magnon states. ${\mathrm{\Gamma }}_2\left(\boldsymbol{q},\boldsymbol{k}\right)$ involves the creation of three bosons, often referred to as a \textquotedblleft{source vertex}\textquotedblright. For the explicit expression, refer to Chernyshev et al.\cite{PhysRevB.79.144416} This kind of the cubic anharmonicity term plays an important role in the unusual dynamical properties of the magnon spectrum, and the magnon energies are strongly renormalized from their harmonic approximation values. To calculate the $1/S$ correction to the magnon spectra, one can conveniently use the standard Green's function approach. Starting from a bare magnon propagator $G^{-1}_0\left(\boldsymbol{k},\omega \right)\mathrm{=}\omega -{\varepsilon }_{\boldsymbol{k}}+i0$, a lowest-order normal self-energy correction by the cubic vertexes is expressed by
\[
{\mathrm{\Sigma }}_1\left(\boldsymbol{k},\omega \right)=\frac{1}{2}\sum_{\boldsymbol{q}}{\frac{{\left|{\mathrm{\Gamma }}_1\left(\boldsymbol{q}\boldsymbol{;}\boldsymbol{k}\right)\right|}^2}{\omega -{\varepsilon }_{\boldsymbol{q}}-{\varepsilon }_{\boldsymbol{k}\boldsymbol{-}\boldsymbol{q}}+i0}}, \tag{7} \label{self1}
\]
\[{\mathrm{\Sigma }}_2\left(\boldsymbol{k},\omega \right)=\frac{1}{2}\sum_{\boldsymbol{q}}{\frac{{\left|{\mathrm{\Gamma }}_2\left(\boldsymbol{q},\boldsymbol{k}\right)\right|}^2}{\omega +{\varepsilon }_{\boldsymbol{q}}+{\varepsilon }_{\boldsymbol{k}\boldsymbol{+}\boldsymbol{q}}-i0}}. \tag{8} \label{self2}
\] 
The new interacting Green's function can be determined by $G^{-1}\left(\boldsymbol{k},\omega \right)\mathrm{=}\omega -{\overline{\varepsilon }}_{\boldsymbol{k}}-{\mathrm{\Sigma }}_1\left(\boldsymbol{k},\omega \right)-{\mathrm{\Sigma }}_2\left(\boldsymbol{k},\omega \right)$, where ${\overline{\varepsilon }}_{\boldsymbol{k}}\mathrm{=}{\varepsilon }_{\boldsymbol{k}}+\delta {\varepsilon }^{(4)}_{\boldsymbol{k}}$ and $\delta {\varepsilon }^{(4)}_{\boldsymbol{k}}$ denotes the contribution of the quartic correction terms by the Hartree-Fock decoupling,\cite{PhysRevB.3.961,PhysRev.117.117} explicitly expressed in Chernyshev et al.\cite{PhysRevB.79.144416}

Such renormalization already gives rise to interesting features in the magnon spectrum: e.g., a strong momentum dependence, renormalized magnon energies being flattened around the M point, a local minimum at the M point (see Fig. \ref{f2}(a)). The amount of renormalization depends on the coupling amplitude between the one-magnon branch and the two-magnon continuum. Interestingly, the feature arising from this renormalization resembles the characteristic minimum of the roton excitation in ${}^{4}$He, and thus it is sometimes called as a \textquotedblleft{roton-like minimum}\textquotedblright. Second, single magnons can decay into two-magnon states if the following conditions are satisfied: (1) ${\mathrm{\Gamma }}_1\left(\boldsymbol{q}\boldsymbol{;}\boldsymbol{k}\right)$, the decay vertex function is non-zero, and (2) the kinematic conditions of momentum and energy conservation are satisfied. Inside the two-magnon continuum, single-magnons are subject to considerable damping. This then leads to the significant broadening of the one-magnon peaks, shown by the gray areas in Fig. \ref{f2}(a). Such magnon damping effect is usually captured in first approximation by the following decay rate, 
\[
{\mathrm{\Gamma }}_{\boldsymbol{k}}\mathrm{=}\frac{\pi }{2}\sum_{\boldsymbol{q}}{{\left|{\mathrm{\Gamma }}_1\left(\boldsymbol{q};\boldsymbol{k}\right)\right|}^2\delta \left({\varepsilon }_{\boldsymbol{k}}-{\varepsilon }_{\boldsymbol{q}}-{\varepsilon }_{\boldsymbol{k}\boldsymbol{-}\boldsymbol{q}}\right)}. \tag{9} \label{decay_rate}
\] 
Fig. \ref{f2}(b) shows the decay rate for the $S$ = 1/2 TLAF. One remarkable feature here is that there are logarithmic singularities at several k points, indicated as \textbf{k}* in the figure. It is naturally arising from the 2D van Hove singularity in the density of states (DOS) of two-magnon continuum. We note that in the higher dimensional case, the DOS of the two-magnon continuum does not exhibit this singularity. Therefore, low-dimensionality plays a crucial role in the magnon-magnon interaction.

It is also worth mentioning the dependence of the magnon-magnon interaction on the spin value. Mourigal et al.\cite{PhysRevB.88.094407,PhysRevB.93.099901} calculated the dynamical structure factor, the quantity directly related to the INS cross section, for both $S$ = 1/2 and $S$ = 3/2. They found that both cases have all the predicted general features: renormalization of the magnon spectrum and broadening of quasiparticle peaks, accompanied by non-Lorentzian features. These effects are significantly stronger in the $S$ = 1/2 case, as shown in Fig. \ref{f2}(d,e). It is also important to note that the spectral weight transfer from the one-magnon to two-magnon continuum is stronger for the $S$ = 1/2 case. 

\subsubsection{$XXZ$ model}
\label{sec:2.1.6}
It is crucial to account for the effect of anisotropy in the Hamiltonian on the magnon decay.\cite{PhysRevB.79.144416} The $XXZ$ spin Hamiltonian is given by
\[
{\mathcal{H}}_{XXZ}\mathrm{=}J\sum_{\left\langle ij\right\rangle }{\left[S^x_iS^x_j+S^y_iS^y_j+\Delta S^z_iS^z_j\right]}, \tag{10} \label{XXZ}
\] 
where $\Delta$ is the two-ion anisotropy parameter. Note that $\Delta<1$ stabilizes the 120$^\circ$ magnetic order as in the isotropic case and the magnon energy in the LWST regime has the slightly modified form of ${\varepsilon }_{\boldsymbol{k}}$ as discussed in Sect. \ref{sec:2.1.4}: ${\varepsilon }_{\boldsymbol{k}}=3JS\sqrt{\left(1-{\gamma }_{\boldsymbol{k}}\right)\left(1+2\Delta {\gamma }_{\boldsymbol{k}}\right)}$. We comment that the higher order terms also keep the more or less same formula, only with minor changes. Fig. \ref{f2}(c) shows the comparison of decay region (shaded area) in the case of the Heisenberg model $\left(\Delta=1\right)$ and $XXZ$ model $\left(\Delta<1\right)$. With decreasing $\Delta$, the decay region (shaded area) becomes smaller and a hole of forbidden decay region develops in the center of the Brillouin zone for $\Delta=0.96$ (See Fig. \ref{f2}(c)). This means that the spontaneous decay process gets markedly suppressed for the anisotropic cases.

\subsection{Magnon-phonon coupling in TLAF}
\label{sec:2.2}
Now let us turn to the interaction between magnons and phonons in the TLAF. Magneto-elastic coupling is a well-known concept in condensed matter physics with a long history. The static properties of such coupling, the spin-lattice coupling, were extensively studied via several experimental techniques. For example, abrupt changes in macroscopic properties such as lattice parameter (magneto-striction),\cite{PhysRevB.5.3642,PhysRev.138.A507,PhysRev.138.A216,PhysRev.96.302} elastic constants,\cite{PhysRevB.83.054418,PhysRevB.76.174426,PhysRevB.10.186,PhysRev.129.1063} and spectroscopic measurements (Raman, IR)\cite{PhysRevB.77.092412,APL.90.151914,PhysRevB.73.214301,PhysRevB.69.214428,JAP.64.5876,PhysRev.127.432} have been observed for several systems. Of further interest, the dynamical aspect of the spin-lattice coupling, i.e. a magnon-phonon coupling, was extensively studied to explain the gap opening between magnon and acoustic phonon branches in TMX${}_{2}$-type collinear magnet compounds (TM=$3d$ transition metal, X=halogen)\cite{PhysRevB.8.2130,J.Phys.C.Solid.6.3156,J.PhysC.5.2769,PhysRevB.32.182} and rare-earth metals.\cite{PhysRevB.12.320,J.Phys.C.Solid.9.111} More recently, a spin-lattice coupling in multiferroic materials was actively studied in the context of magneto-electric coupling.\cite{PhysRevB.88.060103,ActaCrysB.72.3} Thanks to improvements in both the experimental and theoretical tools, it has become possible to carry out in-depth investigations of the dynamical effects of magnon-phonon hybridization, which cannot be fully understood within the spin wave theory. In particular, noncollinear antiferromagnets have attracted much attention due to the sizable hybridization effects in their magnon band structure. In the following subsections, several mechanisms of magnon-phonon coupling shall be introduced. In addition, we will clarify the main effects on magnons and phonons due to the interactions between them, especially for the 2D TLAF with the 120$^\circ$ magnetic order.

\begin{figure}
	\includegraphics[width=1\columnwidth,clip]{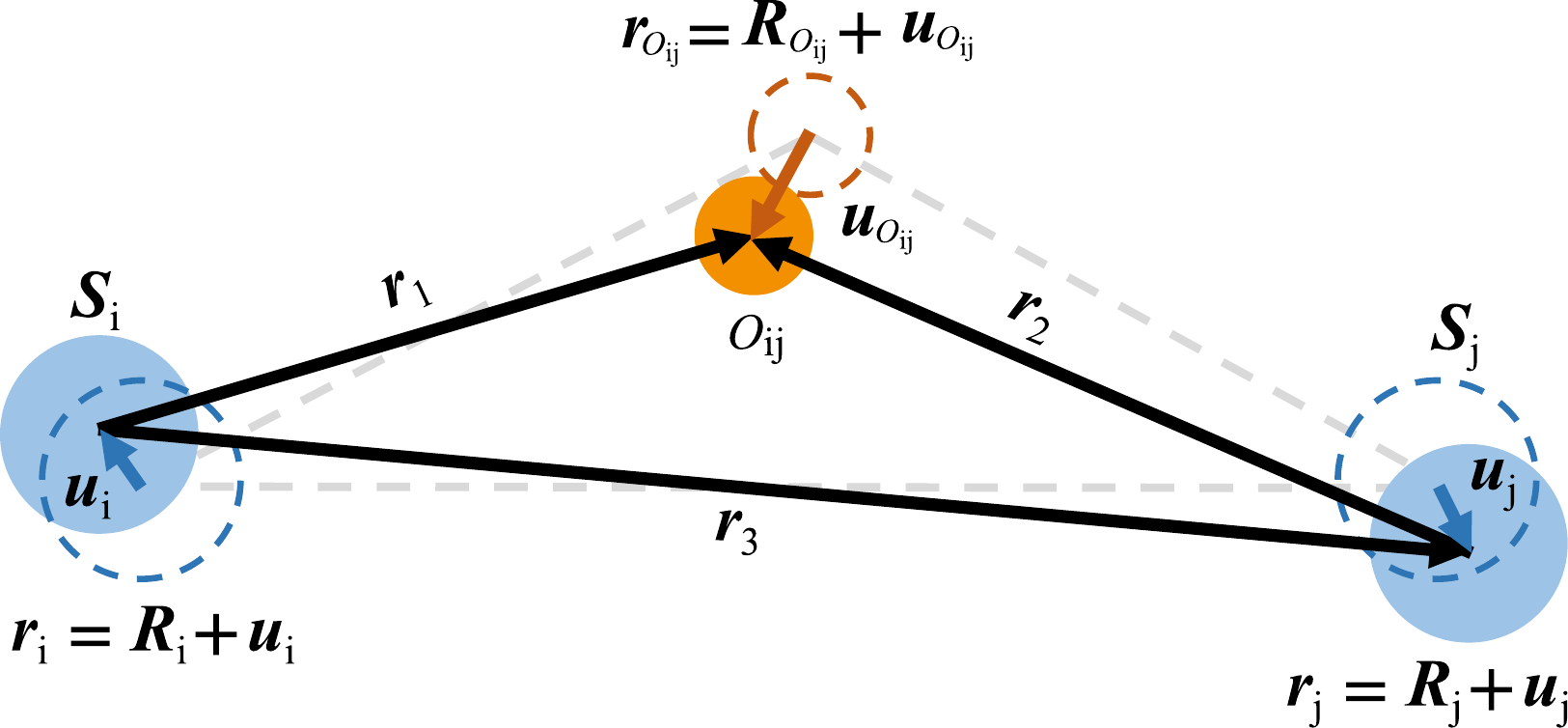}
	\caption{Schematic diagrams for the exchange interactions and relative motions of each atom.}
	\label{f3}
\end{figure}

\subsubsection{Exchange-striction}
\label{sec:2.2.1}
In this subsection, we would like to highlight one of the extensively studied mechanisms, which is supposed to facilitate a magnon-phonon coupling for most magnetic systems: exchange-striction, where exchange energies are modulated as a function of bond vectors of magnetic atoms.\cite{PhysRevB.71.212406,PhysRevB.76.054431,PhysRevB.74.134409} Generally speaking, exchange interaction $J_{ij}$ depends on $r_{ij}=\left|{\boldsymbol{r}}_j-{\boldsymbol{r}}_i\right|$, the distances between spins ${\boldsymbol{S}}_i$ and ${\boldsymbol{S}}_j$ at ${\boldsymbol{r}}_i$ and ${\boldsymbol{r}}_j$. Following the empirical power law: $J_{ij}\propto {1}/{r^n_{ij}}$, it therefore leads to the magnon-phonon coupling constant, which is an index for a magnon-phonon coupling strength, $\mathrm{\alphaup }^\prime\mathrm{=}\partial{J}/\partial{r}\propto n\mathrm{\ \sim \ 8-10}$ for the $3d$ transition metal compounds.\cite{SandorThesis,PhysRevB.44.4657,NatCommun.7.13146,PhysRevB.97.201113,PhysRevB.94.104421} The exchange-striction coefficient, which is a dimensionless unit of the magnon-phonon coupling constant, is expressed as $\alpha=\frac{d}{J}\alpha^\prime$, where $d$ is the half of the bond length and $J$ is the exchange interaction. In many transition metal oxides, the main mechanism of exchange interaction is of super-exchange type, the indirect exchange interaction mediated via non-magnetic atoms (e.g. oxygen). Further consideration of the relative motion of these intermediate atoms with regard to the magnetic atoms also produces a considerable contribution to the exchange-striction mechanism.\cite{AnnalenDerPhys.523.995,SolidStateCommun.149.1749}

The exchange-striction mechanism naturally introduces phonon operators to the magnetic Hamiltonian. It is important to note here that whilst this mechanism does not permit a linear coupling between magnons and phonons in collinear magnets, the linear coupling is in principle allowed for noncollinear spin orders. This is because a magnon operator term ${\mathcal{H}}_1$ originating from the mixing between transverse and longitudinal fluctuations can be coupled to phonon operators via the displacement vectors. The displacement vector of the $i$$^{th}$ atom at ${\boldsymbol{R}}_i$ is given by
\[
{\boldsymbol{u}}_i=\sum_{\boldsymbol{k},\lambda}{\sqrt{\frac{\hslash }{2Nm_i{\omega }_{\boldsymbol{k},\lambda }}}{\boldsymbol{e}}_{\boldsymbol{k},\boldsymbol{\lambda }}\left(b_{\boldsymbol{k},\boldsymbol{\lambda }}+b^{\dagger }_{\boldsymbol{-}\boldsymbol{k},\boldsymbol{\lambda }}\right)e^{i\boldsymbol{k}\cdot {\boldsymbol{R}}_i}},  \tag{11} \label{disp_vec}
\] 
where $N$ is the number of the unit cells, $m_i$ is the atomic mass of $i$$^{th}$ atom, ${\omega }_{\boldsymbol{k},\lambda }$ and ${\boldsymbol{e}}_{\boldsymbol{k},\boldsymbol{\lambda }}$ are the energy and eigenvector of $\lambda$$^{th}$ phonon branch, and $b_{\boldsymbol{k},\boldsymbol{\lambda }}$ ($b^{\dagger }_{\boldsymbol{k},\boldsymbol{\lambda }})$ denotes the annihilation (creation) operator of $\lambda$$^{th}$ phonon branch. It can in principle lead to non-negligible hybridization effects on the magnon and phonon spectra, making the TLAF specially a good model system for the study of the magnon-phonon coupling. In addition, further expansion makes possible the otherwise forbidden spontaneous decay of the hybridized magneto-elastic excitation: one-phonon--two-magnon decay and one-magnon--two-phonon decay. 

Now we would like to introduce the most generalized form of the exchange-striction terms using a Taylor-series expansion around the equilibrium positions in powers of the difference vectors between three atoms,
\begin{align*}
J_{ij}\left({\boldsymbol{r}}_i,~{\boldsymbol{r}}_j,{\boldsymbol{r}}_{O_{ij}}\right)&=J_{ij}\left({\boldsymbol{R}}_i,~{\boldsymbol{R}}_j,\ {\boldsymbol{R}}_{O_{ij}}\right)\\
&\mathrm{+}\left({\boldsymbol{u}}_1\cdot {\nabla }_1+{\boldsymbol{u}}_2\cdot {\nabla }_2+{\boldsymbol{u}}_3\cdot {\nabla }_3\right)J_{ij}  \tag{12} \label{Talyor-expan}\\
&+\frac{1}{2!}{\left({\boldsymbol{u}}_1\cdot {\nabla }_1+{\boldsymbol{u}}_2\cdot {\nabla }_2+{\boldsymbol{u}}_3\cdot {\nabla }_3\right)}^2J_{ij}+\dots,
\end{align*}
where ${\boldsymbol{r}}_i={\boldsymbol{R}}_i+{\boldsymbol{u}}_i$ is the true position vector that deviates from the equilibrium position ${\boldsymbol{R}}_i$ by the phonon displacement vector ${\boldsymbol{u}}_i$. As in Fig. \ref{f3}, we define the difference vector as follows: ${\boldsymbol{r}}_1={\boldsymbol{r}}_{O_{ij}}-{\boldsymbol{r}}_i$, ${\boldsymbol{r}}_2={\boldsymbol{r}}_{O_{ij}}-{\boldsymbol{r}}_j$, and ${\boldsymbol{r}}_3={\boldsymbol{r}}_j-{\boldsymbol{r}}_i={\boldsymbol{r}}_1-{\boldsymbol{r}}_2$. Rewriting the first-order expansion term, we can obtain the following form;
\begin{align*}
&\left({\boldsymbol{u}}_1\cdot {\nabla }_1+{\boldsymbol{u}}_2\cdot {\nabla }_2+{\boldsymbol{u}}_3\cdot {\nabla }_3\right)J_{ij}=\left({\boldsymbol{u}}_{O_{ij}}-{\boldsymbol{u}}_i\right)\cdot {\nabla }_1\left(J_{ij}\right)\\
&+\left({\boldsymbol{u}}_{O_{ij}}-{\boldsymbol{u}}_j\right)\cdot {\nabla }_2\left(J_{ij}\right)+\left({\boldsymbol{u}}_j-{\boldsymbol{u}}_i\right)\cdot {\nabla }_3\left(J_{ij}\right). \tag{13} \label{1st-order_expan}
\end{align*}
The first two terms on the right-hand side correspond to the modulation of superexchange interaction, i.e. superexchange-striction. The third term denotes the common magnon-phonon coupling determined by the relative distance between two spins, i.e. direct exchange-striction. These two different types of exchange-striction effects can be, in principle, present for every magnetic material, depending on the characteristic of the dominant exchange interaction. In the case of direct exchange-striction, as mentioned earlier only the modulation of bond length $r_{ij}$ is important. On the other hand, for the super-exchange-striction there are two key motions contributing more than anything else to the amplitude of the interaction: bond length and bond angle change.

\subsubsection{Antisymmetric exchange-striction}
\label{sec:2.2.2}
A similar approach can be applied to the antisymmetric interaction as well. If the local inversion symmetry is broken in the middle of two spins, the following Dzyaloshinskii-Moriya (DM) interaction\cite{PhysRev.120.91,J.Phys.Chem.Solid.4.241} can be added to the spin Hamiltonian:
\[
{\mathcal{H}}_{spin}\mathrm{=}{\mathcal{H}}_{Heis}+{\mathcal{H}}_{DM}, 
\] 
\[
{\mathcal{H}}_{DM}=\sum_{\left\langle ij\right\rangle }{{\boldsymbol{D}}_{ij}\cdot {\boldsymbol{S}}_i\times {\boldsymbol{S}}_j},  \tag{14} \label{DM}
\] 
where ${\boldsymbol{D}}_{ij}$ is a local DM vector determined by the positions of off-centered intermediate (non-magnetic) atoms. Typically, the amplitude of DM interaction is not nearly as strong, when compared to the exchange interaction in most of $3d$ transition metal compounds. A larger DM interaction may be expected in $4d$ or $5d$ transition metal compounds, due to the larger spin-orbit coupling. In this case, the coupling strengths can be of similar magnitude for both the DM interactions and the exchange interactions.

\begin{table*}[t!]
	\caption{{Summary table for magnon-magnon and magnon-phonon interaction formalism for 2D TLAF}}
	\label{t1}
	\begin{center}
		\begin{tabular}{cccc}
			\hline
			\multicolumn{1}{c}{Mechanism} & \multicolumn{1}{c}{Type} & \multicolumn{1}{c}{Coupling Hamiltonian} &  \multicolumn{1}{c}{References} \\
			\hline
			\multirow{3}*{magnon-magnon} & \multirow{2}*{3-magnon} & \multirow{2}*{$H_{3mag}$ = $\frac{1}{2!}$$\sum_{\boldsymbol{k},\boldsymbol{q}}{\Gamma_1\left(\boldsymbol{q;k}\right)\left(\alpha^\dagger_{\boldsymbol{k-q}}\alpha^\dagger_{\boldsymbol{q}}\alpha^{}_{\boldsymbol{k}}+h.c.\right)}$} & {Eq.\eqref{H3_alpha}} \\&&& {Chernyshev et al.\cite{PhysRevB.79.144416}}\\
			& Predictions: & \multicolumn{2}{c}{spontaneous two-magnon decays, energy renormalization, singularities in the decay rate}\\
			\hline
			\multirow{9}*{exchange-striction} & \multirow{2}*{1-magnon--1-phonon} & \multirow{2}*{$H_{1mag}^{1pho}$ = $\sum_{\boldsymbol{k},\lambda}{g_{\boldsymbol{k},\lambda}\left(a^\dagger_{\boldsymbol{k}}-a^{}_{\boldsymbol{-k}}\right)b_{\boldsymbol{k},\lambda}+h.c.}$} & {Eqs.\eqref{H1pho1mag} and \eqref{gk}} \\&&& {Oh et al.\cite{NatCommun.7.13146} and T\"oth et al.\cite{NatCommun.7.13547}} \\
			& Predictions: & \multicolumn{2}{c}{energy renormalization, energy level repulsion, emergence of hybridized mode}\\
			\cline{2-4}
			& \multirow{2}*{2-magnon--1-phonon} & \multirow{2}*{$H_{2mag}^{1pho}$ = $\sum_{\boldsymbol{k},\boldsymbol{q}}{\Gamma^{mp}\left(\boldsymbol{q,k-q;k}\right)\left(\alpha^\dagger_{\boldsymbol{q}}\alpha^\dagger_{\boldsymbol{k-q}}b^{}_{\boldsymbol{k}}+h.c.\right)}$} & \multirow{2}*{Oh. et al.\cite{NatCommun.7.13146}} \\&&&\\
			& Predictions: & \multicolumn{2}{c}{spontaenous decays of hybridized modes, intrinsic linewidth broadening due to the decay}\\
			\cline{2-4}
			& \multirow{2}*{ESP model} & \multirow{2}*{$H_{ESP}$ = $-JcS^{-2}\sum_i{{\left(\sum_{j\in N\left(i\right)}{\left(\boldsymbol{S}_i\cdot\boldsymbol{S}_j\right){\hat{e}}_{ij}}\right)}^2}$} & {Eq.\eqref{ESP}} \\&&& {Wang et al.\cite{PhysRevLett.100.077201}}\\
			& Predictions: & \multicolumn{2}{c}{energy renormalization}\\
			\hline
			\multirow{3}*{single-ion magneto-striction} &\multirow{2}*{1-magnon--1-phonon} & \multirow{2}*{$H_{SL}$ = $-\sum_{i,\sigma}{{B}^{\sigma }_i{\varepsilon }^{\sigma }_i{S}^{\sigma }_i}$} & {Eq.\eqref{single-ion_magneto}} \\&&&{Callen et al.\cite{PhysRev.139.A455}}\\
			& Predictions: & \multicolumn{2}{c}{energy renormalization, energy level repulsion, intermixing between magnon and phonon}\\
			\hline
		\end{tabular}
	\end{center}
\end{table*}

\subsubsection{Single-ion magneto-striction}
\label{sec:2.2.4}
Lattice vibrations also affect the atomic environment around magnetic atoms, changing the crystal field splittings.\cite{PhysRevB.71.092405} Therefore, they can modulate the single-ion anisotropy via another mechanism of magneto-elastic coupling: single-ion magneto-striction. The concept of such spin-lattice coupling was first introduced by Van Vleck.\cite{PhysRev.57.426} The subsequent studies by Callen et al.\cite{PhysRev.129.578,PhysRev.139.A455} considered the crystal symmetries to formulate a theory of single- and two- ion magneto-striction for ferromagnetic cubic and hexagonal structures. According to their works, the single-ion magneto-striction type coupling Hamiltonian in hexagonal lattice can be defined as
\[
{\mathcal{H}}_{SL}=-\sum_i{\sum_{\sigma }{{B}^{\sigma }_i{\varepsilon }^{\sigma }_i{S}^{\sigma }_i}},  \tag{15} \label{single-ion_magneto}
\] 
where ${B}^{\sigma }_i$ are the magneto-striction coefficients, ${\varepsilon }^{\sigma }_i$ and ${S}^{\sigma }_i$ denote the symmetric representation of the strain tensor and spin operators in terms of irreducible representations $\sigma $  in hexagonal symmetry. For explicit expressions, see Callen et al.\cite{PhysRev.139.A455} As the strain tensor is defined by ${\varepsilon }^{\alpha \beta }_i=\frac{1}{2}\left[\frac{\partial u^{\alpha }_i}{\partial \beta }+\frac{\partial u^{\beta }_i}{\partial \alpha }\right]$ in Cartesian coordinates, Eq. \eqref{single-ion_magneto} includes one-phonon operator and magnon operator of single atom, producing magneto-elastic contributions to the spin wave energies. This single-ion magneto-striction was proposed as an explanation for the gap opening between magnon and acoustic phonon branches that were observed around the $\mathrm{\Gamma }$ point for several rare-earth metals. However, in $3d$ transition metal compounds, this single-ion exchange-striction is usually considered to be smaller than the exchange-striction term.

\subsubsection{Quadratic terms: hybridization of one-magnon and one-phonon}
\label{sec:2.2.5}
In the following subsections, it is assumed for simplicity that the nearest-neighbor exchange coupling is given by a function of $r_{ij}$, an inter-ion distance. Thus, only one coupling constant is needed for the discussion of magnon-phonon coupling. Using Eqs. \eqref{Heis} and \eqref{Talyor-expan}, Heisenberg Hamiltonian can be expanded as follows:
\[
{\mathcal{H}}_{Heis}\mathrm{=}\sum_{\left\langle ij\right\rangle }{J_0\left(1-\frac{\alpha^\prime }{J_0}{\hat{e}}_{ij}\cdot \left({\boldsymbol{u}}_j-{\boldsymbol{u}}_i\right)\right){\boldsymbol{S}}_i\cdot {\boldsymbol{S}}_j},  \tag{16} \label{Heis_alpha}
\] 
where $J_0$ is the exchange interaction when all atoms are at their equilibrium position, ${\hat{e}}_{ij}$ denotes a unit vector connecting magnetic ions at ${\boldsymbol{r}}_i$ and ${\boldsymbol{r}}_j$, and $\alpha^\prime$ is a magnon-phonon coupling constant. Then the generalized Hamiltonian of quadratic order contains the following terms;
\[
{\mathcal{H}}_2^{mp}={\mathcal{H}}_{2mag}+{\mathcal{H}}^{1pho}_{1mag}\mathrm{+}{\mathcal{H}}^{2pho},  \tag{17} \label{H2mp}
\] 
where ${\mathcal{H}}_{2mag}=J_0\sum_{\left\langle ij\right\rangle }{{\boldsymbol{S}}_i\cdot {\boldsymbol{S}}_j}$ is the unperturbed Heisenberg Hamiltonian, and ${\mathcal{H}}^{2pho}=\sum_{\boldsymbol{k}}{{\omega }_{\boldsymbol{k},\lambda }\left(b^{\dagger }_{\boldsymbol{k},\lambda }b_{\boldsymbol{k},\lambda }+\frac{1}{2}\right)}$ denotes the non-interacting (harmonic) phonon Hamiltonian. Using Fourier and Bogoliubov transformations as introduced in Sect. \ref{sec:2.1.4}, the magnon-phonon coupling Hamiltonian ${\mathcal{H}}^{1pho}_{1mag}$ can be derived\cite{NatCommun.7.13547}
\[
{\mathcal{H}}^{1pho}_{1mag}=\sum_{\boldsymbol{k}}{{g }_{\boldsymbol{k},\lambda }\left(a^{\dagger }_{\boldsymbol{k}}-a_{\boldsymbol{-}\boldsymbol{k}}\right)b_{\boldsymbol{k},\lambda }}+\mathrm{h}\mathrm{.c.}, \tag{18} \label{H1pho1mag}
\] 
where the coupling amplitude ${g }_{\boldsymbol{k},\lambda }$ is given by
\[
{g }_{\boldsymbol{k},\lambda }\mathrm{=-}i\frac{3}{4}\alpha \sqrt{\frac{S^3\hslash }{M{\omega }_{\boldsymbol{k},\lambda }}}{\boldsymbol{e}}_{\boldsymbol{k},\boldsymbol{\lambda }}\cdot {f}_{\boldsymbol{k}},  \tag{19} \label{gk}
\]
with the geometrical factor ${\boldsymbol{f}}_{\boldsymbol{k}}$ $\mathrm{=}$ $\sum_{\boldsymbol{\delta }}$ ${\mathrm{sin} \left(\boldsymbol{Q}\cdot\boldsymbol{\delta }\right)\ }$ $\left[{\mathrm{cos} \left(\boldsymbol{k}\cdot\boldsymbol{\delta}\right)\boldsymbol{-}\boldsymbol{1}\ }\right]$$\widehat{\boldsymbol{\delta }}$. This way, ${\mathcal{H}}^{mp}_2$ can be rearranged into the following compact form;
\[
{\mathcal{H}}^{mp}_2=\sum_{\boldsymbol{k}}{X^T_{\boldsymbol{k}}\mathcal{H}X_{\boldsymbol{k}}}.  \tag{20} \label{H2mp_re}
\] 
with
\[
X_{\boldsymbol{k}}={\left(a^{\dagger }_{\boldsymbol{k}}\ a_{\boldsymbol{-}\boldsymbol{k}}\ b^{\dagger }_{\boldsymbol{k},\lambda }\ b_{\boldsymbol{-}\boldsymbol{k},\lambda }\right)}^T.  \tag{21} \label{Xk}
\] 
\[
\mathcal{H}=\left( \begin{array}{cccc}
A_{\boldsymbol{k}} & B_{\boldsymbol{k}} & {g }^{\dagger }_{\boldsymbol{k},\lambda } & {g }_{\boldsymbol{k},\lambda } \\ 
B_{\boldsymbol{k}} & A_{\boldsymbol{k}} & {g }_{\boldsymbol{k},\lambda } & {g }^{\dagger }_{\boldsymbol{k},\lambda } \\ 
{g }_{\boldsymbol{k},\lambda } & {g }^{\dagger }_{\boldsymbol{k},\lambda } & {\mathrm{\omega }}_{\boldsymbol{k},\lambda}/2 & 0 \\ 
{g }^{\dagger }_{\boldsymbol{k},\lambda } & {g }_{\boldsymbol{k},\lambda } & 0 & {\mathrm{\omega }}_{\boldsymbol{k},\lambda}/2 \end{array}  \tag{22} \label{H2mp_matrix}
\right)
\] 
${\mathcal{H}}^{1pho}_{1mag}$ contributes to the off-diagonal component in the quadratic form $\mathcal{H}$.\cite{PhysRev.139.A450,NatCommun.7.13146} Note that the Hamiltonian is still Hermitian and $\omega_{\boldsymbol{k},\lambda}/2$ is the phonon energy. Further Bogoliubov transformation yields the energies of the hybridized modes. There are two important features to be noted about the hybridized spectra. First is the level repulsion between the magnon and phonon modes.\cite{PhysRevB.94.104421} This is naturally attributed to the non-zero off-diagonal components of the Hamiltonian matrix. Such repulsion is maximized when the magnon and phonon branches are overlapped, opening a gap between them. A second important feature of the magnon-phonon coupling is the intermixing of the spectral weight.\cite{NatCommun.7.13146} In other words, the spectral weights of magnon or phonon branches can be transferred to one another. This intermixing generates new observable hybridized magneto-elastic excitations. For example, an optical phonon branch near a magnon branch can be detected using INS experiments in the low-momentum region if there is a strong coupling between them.

\subsubsection{Cubic terms: spontaneous decays of hybridized modes}
\label{sec:2.2.6}
Now let us re-examine the cubic anharmonicity term for 2D TLAF by including the phonon contribution. The earlier studies on the nonlinear magento-elastic coupling were focused on the contribution to the transport properties such as the thermal conductivity\cite{PhysRevB.8.2130,Prog.Theor.Phys.29.801,J.Phys.Colloq.42.262,J.PhysC.14.2147} and the acoustic sound wave velocity in the long wavelength limit. In the past decades, there have been a few new theoretical studies on the dynamical aspects of the nonlinear correction (e.g. renormalization and damping of magnons and phonons) based on the Green's function method,\cite{AnnalenDerPhys.523.995,PhysRevB.89.184413,PhysRevB.95.064410,PhysRevB.65.014409,PhysRevB.76.104407} while Kreisel et al.\cite{PhysRevB.84.024414} carried out the detailed application specific to the 2D TLAF with the noncollinear magnetic order. Even though they used a spiral magnetic order with field-induced canting in a distorted triangular lattice, it can in principle be extended to an ideal case. The most general form of the cubic term ${\mathcal{H}}_3^{mp}$ is given by
\[
{\mathcal{H}}_3^{mp}={\mathcal{H}}_{3mag}+{\mathcal{H}}^{1pho}_{2mag}+{\mathcal{H}}^{2pho}_{1mag}\mathrm{+}{\mathcal{H}}^{3pho},  \tag{23} \label{H3mp}
\] 
where ${\mathcal{H}}_{3mag}$ is identical with Eq. \eqref{H3_alpha}. ${\mathcal{H}}^{1pho}_{2mag}$ is the one-phonon and two-magnon interaction term. The third ${\mathcal{H}}^{2pho}_{1mag}$ is a coupling term originating from the second-order Taylor expansion in Eq. \eqref{Talyor-expan}. Thus, it is proportional to the second derivative of the exchange interaction. However, in the weak magnon-phonon coupling regime, such higher-order derivatives are often negligible because the lattice modulations are expected to be very small at low temperatures. Thus, ${\mathcal{H}}^{2pho}_{1mag}$ can also be neglected for our purpose as we mainly concern with the dynamical aspect of the coupling at low temperatures. The anharmonic phonon term ${\mathcal{H}}^{3pho}$ involved in the spontaneous phonon decay is also negligible in most cases. Therefore, only ${\mathcal{H}}^{1pho}_{2mag}$ can make a significant contribution to the hybridized spectra. As the generalized form of ${\mathcal{H}}^{1pho}_{2mag}$ was already derived by Kreisel et al.,\cite{PhysRevB.84.024414} the relevant decay channel is the radiation (absorption) of a phonon by two spin waves:\cite{JoosungThesis} ${\mathrm{\Gamma }}^{mp}\left(\boldsymbol{q},\boldsymbol{k}-\boldsymbol{q};\boldsymbol{k}\right){\alpha }^{\dagger }_{\boldsymbol{q}}{\alpha }^{\dagger }_{\boldsymbol{k}-\boldsymbol{q}}b_{\boldsymbol{k}}+h.c$. From the decay process allowed by this vertex function and the kinematic conditions, phonons near the spin wave dispersion can be radiated or absorbed by the two magnons. It is also accompanied by a significant damping effect that can be observed by INS.\cite{NatCommun.7.13146}

\begin{figure*}
	\includegraphics[width=1\textwidth,clip]{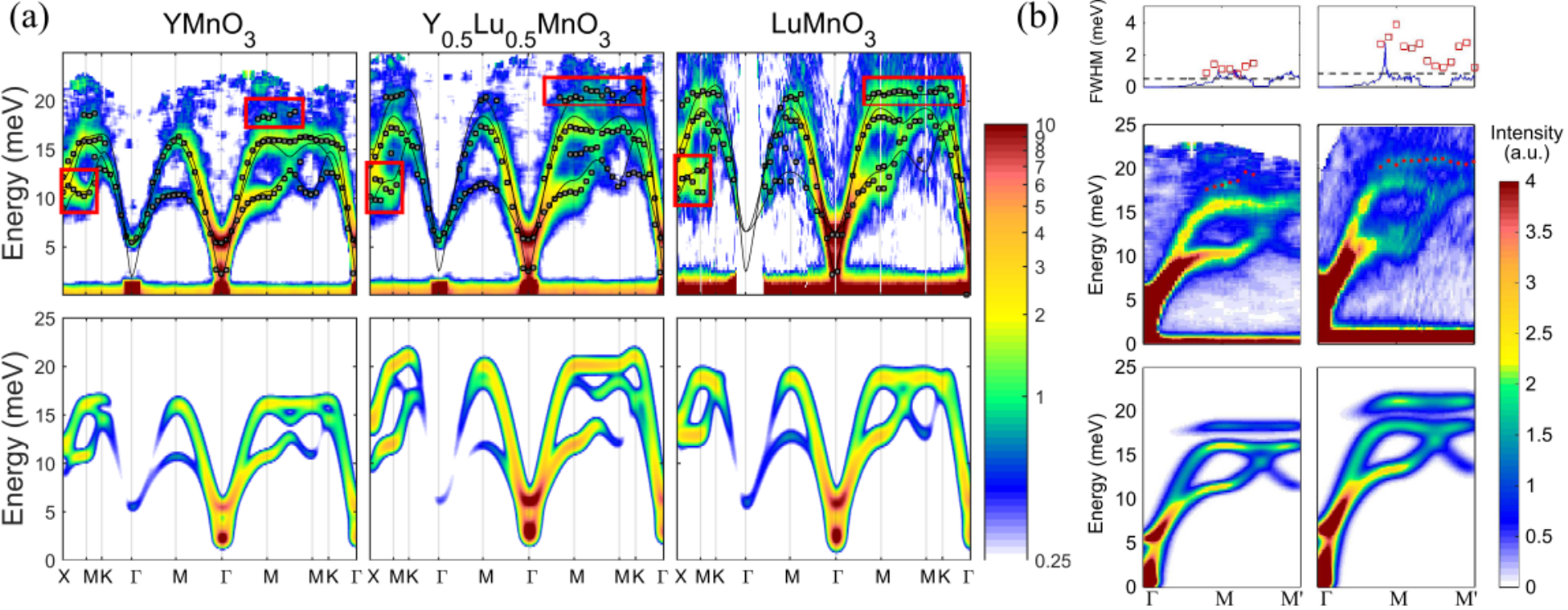}
	\caption{(a) INS data are shown for YMnO$_3$, Y$_{0.5}$Lu$_{0.5}$MnO$_3$, and LuMnO$_3$ in the upper panel. Red boxes indicate the $Q-E$ space, where there are the discrepancies between the experimental data and the LSWT calculation results. The bottom panel displays the calculated spin wave dispersion and the dynamical structure factor using a minimal spin Hamiltonian within the LSWT approximation. (b) Comparison between the experimental data and the magnon-phonon coupling model calculation results for the decay rate and the magneto-elastic excitations. (Upper panel) The full width at half maximum (FWHM) of magnon peaks and the decay rate, (middle panel) the INS data of YMnO$_3$ and LuMnO$_3$ along the $\Gamma$MM$^\prime$ direction, and (bottom panel) the magnon-phonon coupling model calculation. Reprinted with permission from Oh et al.\cite{NatCommun.7.13146} (Copyright © Nature Publishing Group).}
	\label{f4}
\end{figure*}

\subsection{Einstein Site Phonon model}
\label{sec:2.3}
The derivation of the full magnon-phonon coupling Hamiltonian discussed in Sect. \ref{sec:2.2} requires one to introduce the full phonon operators for the given material, which is often technically challenging. However, a simple and yet useful starting point would be to include just one dispersionless optical phonon branch to effectively describe the coupling Hamiltonian. This, so-called Einstein site phonon model,\cite{PhysRevB.74.134409} is based on an assumption of the exchange-striction type spin-phonon coupling. Integrating out the Einstein site phonons in the phonon sector with respect to the atomic displacements, the total Hamiltonian of 2D TLAF can be reduced to the following form;\cite{PhysRevLett.100.077201}
\begin{align*}
\mathcal{H}_{ESP} &=J\sum_{<ij>}\boldsymbol{S}_i\cdot\boldsymbol{S}_j\\
&-JcS^{-2}\sum_i{{\left(\sum_{j\in N\left(i\right)}{\left(\boldsymbol{S}_i\cdot\boldsymbol{S}_j\right){\hat{e}}_{ij}}\right)}^2}. \tag{24} \label{ESP} 
\end{align*}
Here, $c$ is a spin-phonon coupling constant defined as $\mathrm{c=}{\mathrm{\alphaup }}^{\mathrm{2}}JS^2/(2K)$, K is an elastic constant in units of energy, and ${\hat{e}}_{ij}$ is a unit vector that points from site \textit{j} to site \textit{i}. \textit{N(i)} represents the nearest-neighbor sites around the \textit{i}$^{th}$ atom. Note that the magnitude of $c$ is a key parameter in this Hamiltonian, as it is an index of coupling strength and determines the magnetic ground state. For example, in a general phase diagram of the magnetic ground state in 2D TLAF, the 120$^\circ$ spin ordered state is reported to be stable up to $c$ = 1/8.\cite{PhysRevLett.100.077201}

\section{Experimental Observations}
\label{sec:3}
Over the past decades, there has been a slow but steady growth in the volume of experimental studies on the magnon-magnon and magnon-phonon coupling in various magnetic systems. As outlined earlier, our aim for this review is concentrated on the noncollinear magnets that can have three-magnon interactions and linear magnon-phonon coupling terms like 2D TLAF. So far, there are several triangular lattice materials that exhibit a 120$^\circ$ spin ordered state with different spin values:\cite{Can.J.Phys.75.605} RbFe(MoO$_4$)$_2$,\cite{PhysRevB.74.024412,PhysRevLett.98.267205,PhysRevB.67.094434,PhysRevB.88.060409} hexagonal RFeO$_3$\cite{ModernPhysLettB.28.1430008} for $S$ = 5/2, RMnO$_3$\cite{PhysRevB.68.104426,JAP.93.8194,ActaCrysB.72.3} with R=rare-earth elements for $S$ = 2, ACrO$_2$\cite{PhysRevLett.101.067204} with A=Li, Ag, or Cu for $S$ = 3/2, Ba$_3$CoM$_2$O$_9$\cite{PhysRevLett.109.267206,PhysRevB.90.014403,JPCM.29.115804} with M=Sb, Nb, or Ta, CsCuCl$_4$\cite{PhysRevLett.86.1335}, and CsCuBr$_4$\cite{PhysRevLett.102.257201} for $S$ = 1/2. In this section, we focus on the magnetic excitations of three specific cases: the hexagonal rare-earth manganite (h-RMnO$_3$) with R=Y, Lu, or Ho, the hexagonal LuFeO$_3$\cite{PhysRevB.98.134412,PhysRevB.97.184419} (h-LuFeO$_3$), the delafossite compounds (ACrO$_2$), and the $S$ = 1/2 systems of Ba$_3$CoSb$_2$O$_9$ and Cs$_2$CuCl$_4$.

\subsection{Decay and hybridization of magnetic excitations in RMnO$_3$}
\label{sec:3.1}
The crystal structure of h-RMnO$_3$ has the $P6_3cm$ space group, in which MnO$_5$ bipyramids form a triangular lattice in the $ab$ plane. The rare-earth atoms are located between the layers consisting of Mn$^{3+}$ ions. The magnetic structure of h-RMnO$_3$ is known to have the 120$^\circ$ spin ordered state, and can be represented by the four magnetic representations of $\Gamma$$_1$$\sim$$\Gamma$$_4$\cite{JAP.93.8194} and their linear combinations. The dominant exchange interaction between Mn$^{3+}$ ions with $S$ =  2 is a super-exchange type interaction along the Mn-O-Mn bond in the $ab$ plane. Interestingly, Mn trimerization\cite{PhysRevB.86.054407,ActaCrysB.72.3} occurs in RMnO$_3$ so that the strengths of intra- and inter-trimer super-exchange interactions are different, as evidenced in earlier INS experiments.\cite{PhysRevB.82.184420,PhysRevB.68.014432} Except for the trimerization effect, RMnO$_3$ is a good real-world 2D TLAF manifestation for comparison with ideal 2D TLAF models. Importantly, the inter-plane exchange interaction between the Mn adjacent layers is smaller by two orders of magnitude compared with the in-plane one,\cite{PhysRevLett.111.257202} which can be negligible for most of the cases. One more advantage of h-RMnO$_3$, it is reasonably easy to grow high quality single crystals large enough for INS experiments. 

h-RMnO$_3$ is also known as a technologically promising material due to its multiferroic properties.\cite{NatMat.6.13,NatRevMat.1.16046,JKPS.63.504} A coupling between the coexisting ferroelectric and magnetic ordered states could provide novel functionalities or phenomena that would be useful in future applications. Indeed, a giant magneto-elastic coupling\cite{Nature.451.805} and a strong magneto-electric effect were discovered to be present in these types of materials. In this context, one of the key issues is the correlation between spin and lattice degrees of freedom, and thus studying the spin-lattice coupling is important for uncovering the underlying microscopic mechanisms in these intriguing multiferroic materials. As such, the experimental evidence of the spin-lattice coupling has been previously observed in a series of experiments: X-ray and neutron diffraction,\cite{PhysRevB.82.054428,Nature.451.805,PhysRevB.71.180413,JPCM.24.336003,PhysRevB.94.125150} Raman and infrared optical spectroscopy,\cite{JPCM.20.425219,JPCM.22.356002,JPCM.16.809,JAP.121.084102,SolidStateScience.11.1639,JPCM.19.365239,JPCM.21.064218,PhysRevB.56.2488,PhysRevB.90.024307} thermal conductivity,\cite{PhysRevLett.93.177202,PhysRevB.96.174425} elastic moduli,\cite{PhysRevB.83.054418,PhysRevB.76.174426} and thermal expansion\cite{PhysRevB.71.060407,JPCM.29.095602} measurements.

As introduced in Sect. \ref{sec:2.1}, the INS data for LuMnO$_3$ revealed some interesting features that are readily identified with the cubic and higher order terms of the spin Hamiltonian,\cite{PhysRevB.79.144416} i.e. interactions between magnons: (1) a roton-like minimum at the M point, (2) a flattened top mode of spin wave dispersion, and (3) an intrinsic linewidth broadening of the high energy magnon mode.\cite{PhysRevLett.111.257202} However, it was also noted that the downward shift at the M point cannot be fully explained only by the magnon-magnon interactions.

As already discussed in Sect. \ref{sec:2.2}, it is also possible to have a linear magnon-phonon coupling. Oh et al. indeed observed the direct evidence of such magnon-phonon coupling in (Y,Lu)MnO$_3$ using INS.\cite{NatCommun.7.13146} As shown in Fig. \ref{f4}, there exist clear discrepancies between the experimental data and the LSWT calculation results with the minimal spin Hamiltonian including intra- and inter-trimer exchange interactions ($J_1$, $J_2$), and easy-plane and easy-axis anisotropies ($D_1$, $D_2$) without higher-order terms. The key experimental deviations from LSWT can be summarized as follows: (1) the downward curvature along the XM direction, (2) the fitted $J_1/J_2$ ratio was unrealistic given the facts that the actual difference of the Mn trimer lengths is actually small and so DFT calculations only predicts a finite but small difference between the interactions, $J_1/J_2\sim1$,\cite{PhysRevB.86.054407} (3) the calculated dynamical structure factor shows a different intensity ratio between the top and middle modes of the spin waves from the INS data, and (4) a new excitation mode is found around 20 meV, above the single magnon modes at 15 meV. Taken together, all these features cannot be simply explained by the minimal spin Hamiltonian within the LSWT approximation.

The new excitations observed at 20 meV are most probably due to magneto-elastic modes, originating from hybridization between magnons and phonons as described in Sect. \ref{sec:2.2}. Implementing the DFT calculations for phonons, it was possible to develop the total Hamiltonian including all 90 phonon operators, 6 spin operators and their hybridized terms. The magneto-elastic excitations are then produced in the calculated dynamical structure factor, consistent with the experimental observations, and explained the other formerly discrepant features listed above. The exchange-striction coefficient $\alpha$ is estimated to be 8$\sim$10 for the (Y,Lu)MnO$_3$. Interestingly, this value is similar with the one estimated from the relation between the pressure dependence of Neel temperature and lattice constants.\cite{PhysRevLett.98.197203,NatCommun.7.13146,JETP.82.193}

\begin{figure}
	\includegraphics[width=0.9\columnwidth]{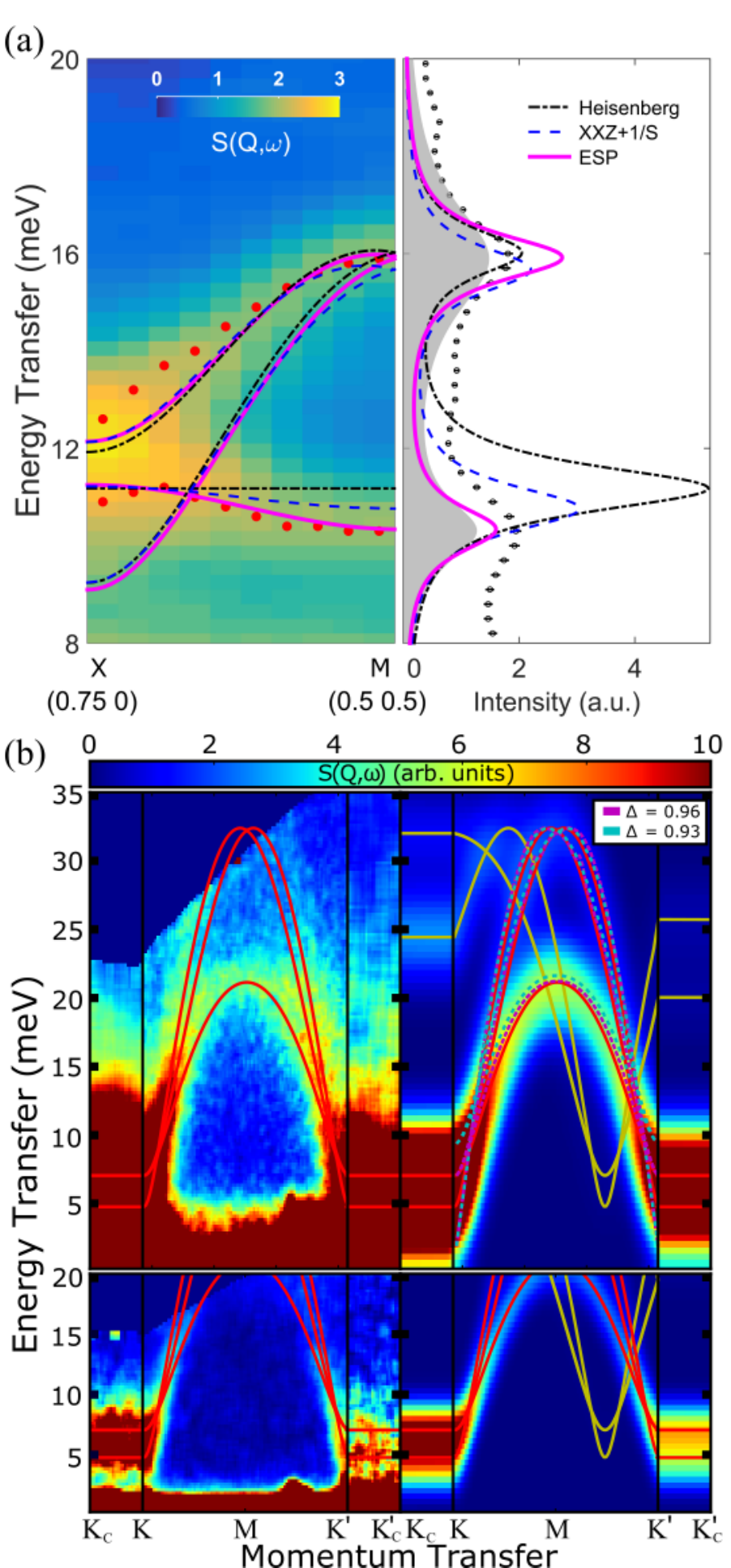}
	\centering
	\caption{(a) INS data for HoMnO$_3$ and comparison with three model calculations: Heisenberg model with LSWT approximation, Heisenberg $XXZ$ model with $1/S$ correction, and the ESP model as indicated in the legend. Right panel shows the intensity comparison between the INS data and the three model calculations at the M point. (b) INS data for Lu$_{0.6}$Sc$_{0.4}$FeO$_3$ and comparison with LSWT calculations and Heisenberg $XXZ$ model with $1/S$ correction. The K$_C$ and K$_C^\prime$ indicate the K points located at different L value with L=1/4. Red (yellow) solid lines represent the LSWT calculation results from the first (second) domain of the crystal. Dashed colored lines represent the $XXZ$ model with $1/S$ correction. Reprinted with permission from Kim et al.\cite{PhysRevB.97.201113} and Leiner et al.\cite{PhysRevB.98.134412} (Copyright © American Physical Society).}
	\label{f5}
\end{figure}

The spontaneous decay of such magneto-elastic excitations is also possible when the kinematic conditions are satisfied. According to the detailed analysis based on the assumption of one dispersionless optical phonon branch,\cite{NatCommun.7.13146} the most appropriate decay process for such  magneto-elastic modes is where they decay into the original lower energy states of the two-magnon continuum. The calculated decay rates in these instances are in good agreement with the measured full width at half maximum (FWHM) of the magneto-elastic modes as shown in upper panel of Fig. \ref{f4}(b).

The magnitude of broadening is an important experimental signature of the magnon decay as it is directly proportional to the number of allowed decay channels. It was experimentally observed that the FWHM of the magneto-elastic mode is significantly larger in LuMnO$_3$ than in YMnO$_3$. This observation indicates that there is a greater number of allowed decay channels in the LuMnO$_3$ case. The reason is believed to be that the magneto-elastic mode for LuMnO$_3$ is found to cross the boundary of two-magnon continuum (for example, at the M point), which allows for its decay in the reciprocal space. Unfortunately, this does not happen in YMnO$_3$ because the magneto-elastic mode does not cross such boundaries and so the spontaneous decay is forbidden for YMnO$_3$. 

Similar features are also found in another hexagonal manganite, HoMnO$_3$, as shown in Fig. \ref{f5}(a).\cite{PhysRevB.97.201113} There is clearly the downward curvature along the XM direction and diffusive signals above the upper magnon branch, which resemble very closely with the magneto-elastic excitation in LuMnO$_3$. Unlike other RMnO$_3$ materials, however HoMnO$_3$ exhibits a very weak Mn trimerization effect, almost negligible at low temperature according to neutron diffraction studies.\cite{PhysRevLett.103.067204} This gives HoMnO$_3$ the advantage of being a nearly ideal 2D TLAF, thus enabling an easy and reliable comparison between theory and experiment. To quantitatively investigate the strength of the magnon-magnon interaction and the magnon-phonon coupling, we used the Heisenberg $XXZ$ model including higher order corrections\cite{PhysRevB.79.144416} and the ESP model\cite{PhysRevLett.100.077201} introduced in Sects. \ref{sec:2.1} and \ref{sec:2.3}. As can be gleaned from the YMnO$_3$ and LuMnO$_3$ cases, the downward curvature along the XM direction may be in principle explained by both models. However, the main difference appears among the three models in the intensity comparison at the M point. For example, the ratio of the intensity for the middle and top modes is only reproduced by the ESP model as shown in Fig. \ref{f5}(a), which includes only a magnon-phonon coupling term. The exchange-striction coefficient $\alpha$ estimated from the ESP model is found to be 12.8, indicating an exceptionally strong magnon-phonon coupling, especially when compared to (Y,Lu)MnO$_3$.

As in YMnO$_3$, the linewidth broadening is not significantly enhanced in HoMnO$_3$. This can be regarded as the consequence of a reduced number of decay channels due to the strong easy-plane anisotropy, which was estimated to be $\Delta$ = 0.88 for HoMnO$_3$. As the easy-plane anisotropy $\Delta$ increases, the boundaries of the decay region in the reciprocal space are drastically reduced and eventually disappear when the anisotropy becomes smaller than the critical value of $\Delta$ = 0.92.\cite{PhysRevB.79.144416} However, it is shown experimentally that the signals of two-magnon continuum state are not completely absent even for HoMnO$_3$ with $\Delta=0.88$, in contrast to the theoretical predictions. But the intensity and linewidth broadening effect are found to be quite small, which indicates the almost complete suppression of magnon decay in HoMnO$_3$. Thus it is concluded that coupling effects acting on the spin wave dispersion and the dynamical structure factor are primarily coming from the magnon-phonon coupling, and the magnon-magnon interaction strength is almost negligibly weak for HoMnO$_3$.

In order to examine the dependence of magnon-phonon coupling on the spin value in 2D TLAF systems, h-LuFeO$_3$ was examined. The Fe$^{3+}$ ions in LuFeO$_3$ have a 3$d^5$ electron configuration ($S$ = 5/2). Recent INS data from a Lu$_{0.6}$Sc$_{0.4}$FeO$_3$ single crystal is reproduced in Fig. \ref{f5}(b).\cite{PhysRevB.98.134412} As one can see, there are no features similar to those expected from both the magnon-phonon coupling and the magnon-magnon interaction such as the roton-like minimum at the M point or flattening of magnon modes. This indicates that there is less overlap between magnon and phonon modes for Lu$_{0.6}$Sc$_{0.4}$FeO$_3$. The other possibility is that the large spin value $S$ =  5/2 of Fe$^{3+}$ may also lead to some suppression of the expected magnon-magnon interaction.\cite{PhysRevB.88.094407}

\begin{figure}
	\includegraphics[width=1\columnwidth,clip]{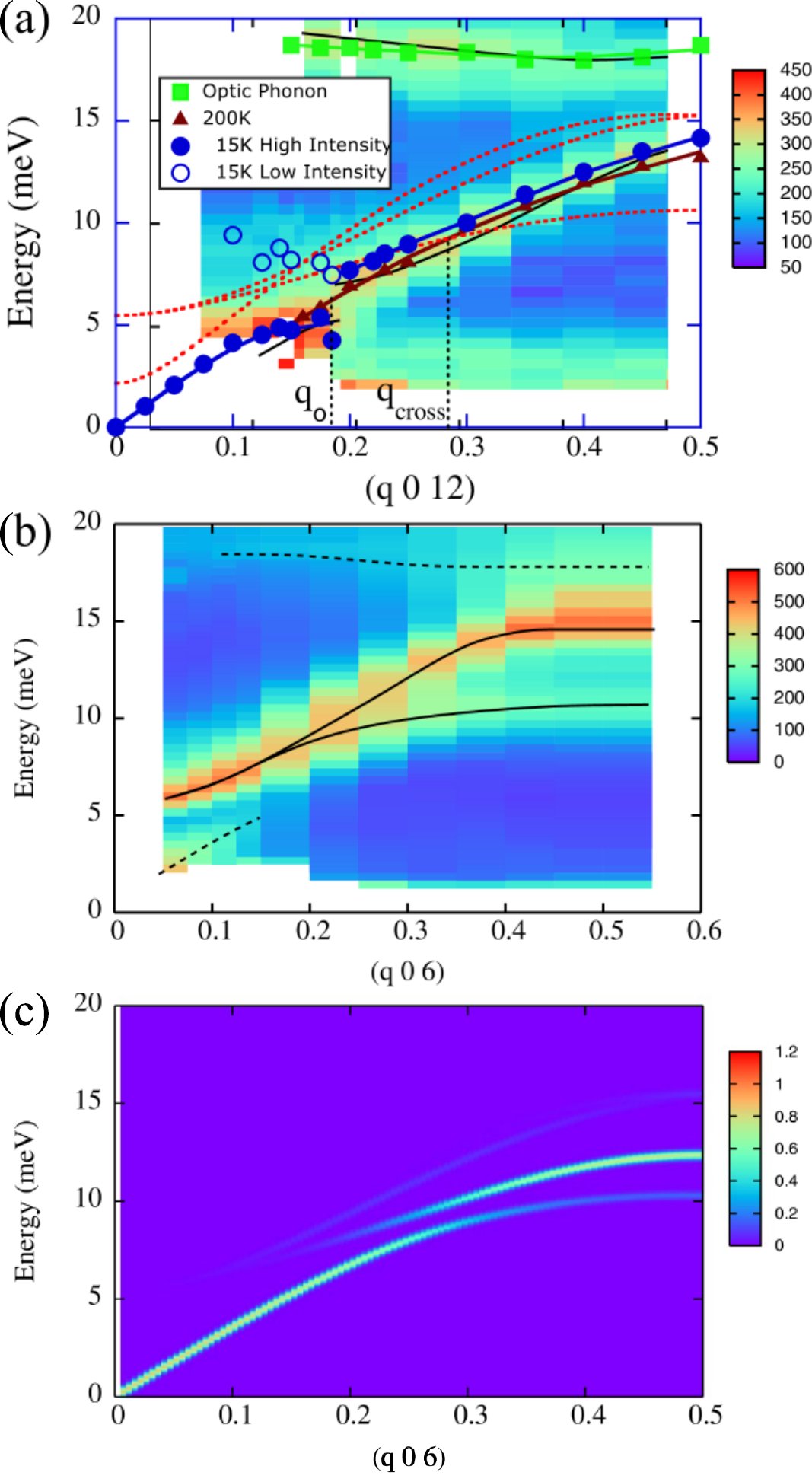}
	\caption{Experimentally measured and theoretically calculated phonon and magnon modes: (a) Circles and triangles denote the observed phonon modes at 18 and 200 K, respectively. Squares show the observed optic phonon modes and dotted lines indicate the calculated magnon modes. A gap is reportedly seen in the acoustic phonon mode in YMnO$_3$ below T$_N$ at high $\vert$$q$$\vert$, whereas (b) no gap is seen in the magnon dispersion down to lower $\vert$$q$$\vert$ values. Solid and dashed lines in (a) and (b) are guides to the eyes. (c) The calculation results are shwon at the same $q$ as in (b) using the single-ion magneto-striction mechanism. Reprinted with permission from Petit et al.\cite{PhysRevLett.99.266604} (Copyright © American Physical Society).}
	\label{f6}
\end{figure}

\begin{figure*}
	\includegraphics[width=1\textwidth,clip]{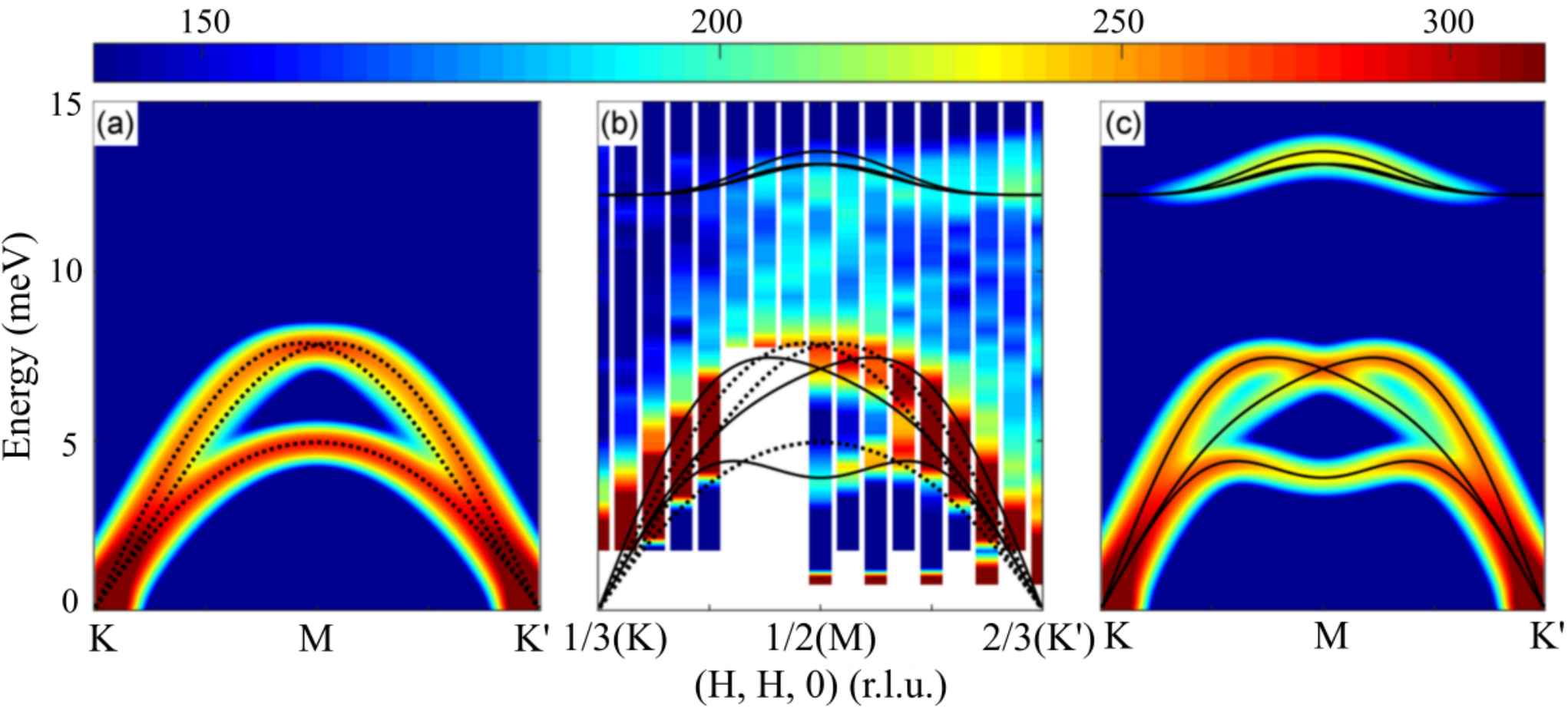}
	\caption{(a) Calculated intensity along the KMK$^\prime$ line for CuCrO$_2$ with LSWT approximation. (b) INS data for CuCrO$_2$. (c) Calculated intensity using a magnon-phonon coupling model. Reprinted with permission from Park et al.\cite{PhysRevB.94.104421} (Copyright © American Physical Society).}
	\label{f7}
\end{figure*}

All discussions on the unconventional magnetic excitations in RMnO$_3$ are so far based on the exchange-striction mechanism introducing a coupling between magnons and phonons. However, we should note that this is not the only way to interpret the experimental observations. For example, Petit et al. reported the opening of a gap in the acoustic phonon branch as shown in Fig. \ref{f6},\cite{PhysRevLett.99.266604} which they interpreted as the result of hybridization between magnon and acoustic phonon with a gap at $q = q_o$ ($\sim$0.185). They assumed in their analysis that the underlying mechanism is a single ion magneto-striction model.\cite{J.PhysC.5.2769} According to their explanations, the vibration of atoms changes the crystal field of MnO$_5$ bipyramids resulting in the modulation of single-ion anisotropy. It also allows spins to couple directly with phonons, which is also mediated by the single-ion anisotropy. Using this mechanism, they successfully reproduced the avoided crossing between the magnon and phonon modes near the K point in the $ab$ plane as discussed in a recent study.\cite{PhysRevB.97.134304} However, we note that the momentum transfer position of $q_o$ ($\sim$0.185), where the gap opens, is not consistent with the experiments: in the actual experimental data, the magnon and acoustic phonon cross each other at $q = q_{cross}$ ($\sim$0.3). Furthermore, the calculation results based on the magneto-striction of single-ion type mechanism failed to explain the observed $q_o$ value.

Another possible mechanism for the hybridization was proposed by Pailhes et al.\cite{PhysRevB.79.134409} Using polarized INS data, they separated the nuclear and magnetic dynamical structure factors of the hybridized modes around $q = q_o$. In their proposed mechanism, the modulation of the DM interaction is considered as an important agency mediating the observed coupling between magnons and phonons. This is in fact the antisymmetric exchange-striction as described in Sect. \ref{sec:2.2}. In a realistic model, the oxygen atoms tend to move spontaneously along the $c$ axis, causing the modulation of the DM interaction. Therefore, the spins are slightly rotated towards the $c$ axis and can be coupled with the motion of oxygen atoms along the $c$ axis. But as far as we can see, it still cannot explain the aforementioned discrepancy in the two different values of the crossing between $q_o$ and $q_{cross}$. Therefore, it still remains not fully explained and warrants further quantitative analysis or the suggestion of a new mechanism. It is also an interesting question how these unusual features in magnons and phonons evolve upon doping.\cite{PhysRevB.79.064417}

\subsection{Strong magnon-phonon coupling in ACrO$_2$ (A = Cu, Li)}
\label{sec:3.2}
The delafossite ACrO$_2$ compounds also form a 2D TLAF with magnetic Cr$^{3+}$ ions of $S$ =  3/2. For the CuCrO$_2$ case, a proper helix magnetic ground state with a propagation vector $Q$ = (0.329, 0.329, 0) could provide the nearly 120$^\circ$ spin ordered state representing the 2D TLAF system.\cite{PhysRevB.81.104411} Since the Cr layers are well separated from each other along the $c$ axis by the nonmagnetic Cu$^+$ ions, the inter-layer coupling is small enough, and can be safely neglected while an in-plane direct exchange interaction between Cr$^{3+}$ ions is dominant. These properties are also similar to the LiCrO$_2$ case, where the magnetic ground state is a helical order in the $ac$ plane with two propagation vectors: $Q$ = (1/3, 1/3, 0) and (-2/3, 1/3, 1/2).\cite{JPCM.7.6869} Interestingly enough, it was discovered that CuCrO$_2$ also exhibits multiferroic behavior.\cite{PhysRevLett.101.067204} At the same time, a slight deformation of the triangular lattice plane,\cite{JPSJ.78.113710} softening of transverse phonon,\cite{PhysRevB.88.224104} and a shift of Raman peaks below T$_N$ were observed for CuCrO$_2$,\cite{JPCM.24.036003} which can be taken as strong evidence for the magneto-elastic coupling and spin-phonon coupling effects in this material.

\begin{figure*}
	\includegraphics[width=1\textwidth,clip]{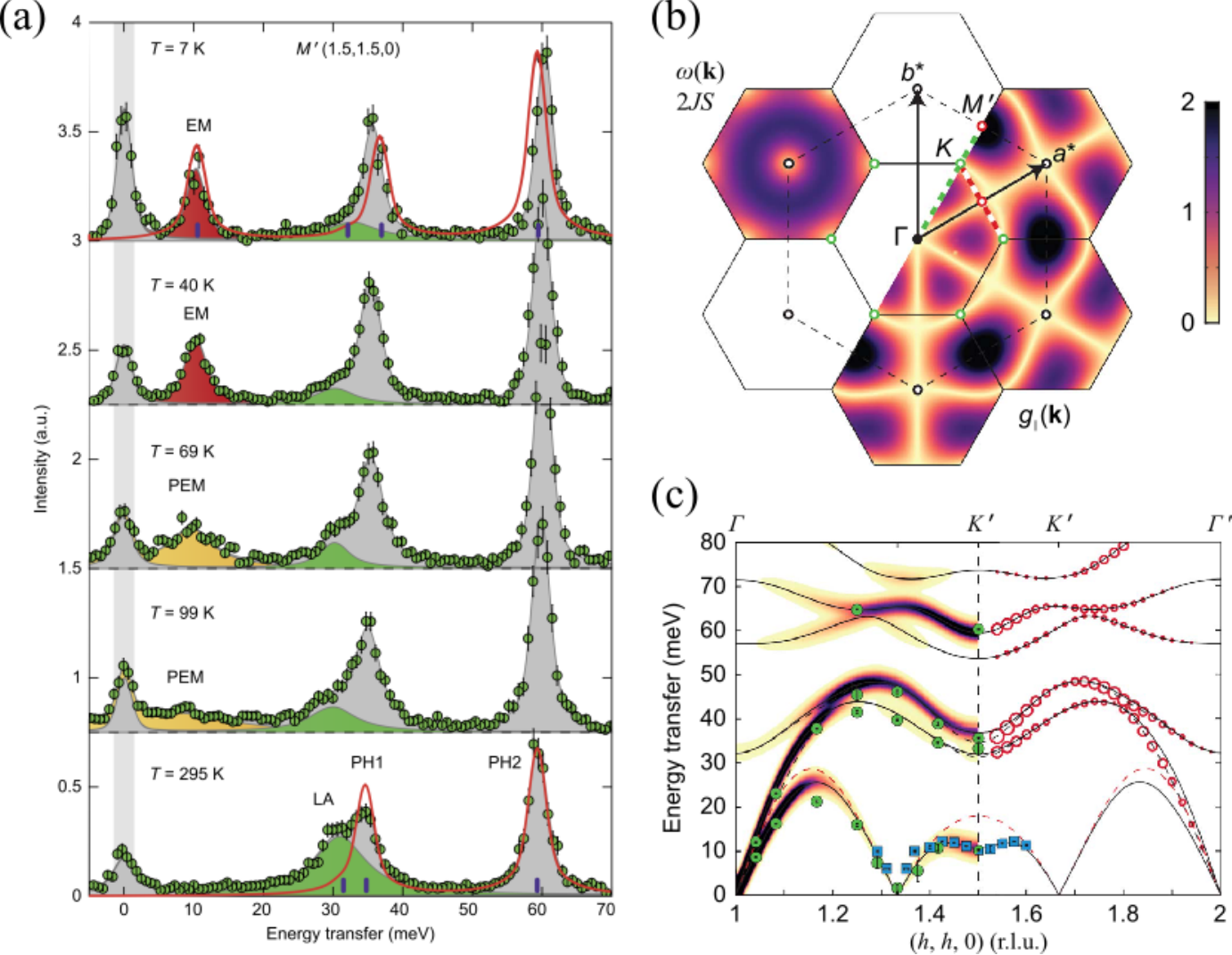}
	\caption{(a) Magnon-phonon hybridized excitations observed by IXS in LiCrO$_2$ due to hybridization between magnons and phonons. (b) Reciprocal space of the triangular lattice with black and dashed hexagons denoting the magnetic and crystallographic Brillouin zones, respectively. (c) Comparison of the measured phonon dispersion at 7 K and the coupled magnon–phonon model along the $(h, h, 0)$ direction. The colour map on the left half shows the calculated IXS cross-section, while the filled green circles and blue squares denote the measured quasiparticle energies using IXS and INS, respectively. The black dashed and red dashed lines indicate the magnon and longitudinal phonon dispersion of the uncoupled model, while the continuous black lines correspond to the coupled dispersion. Reprinted with permission from T\"oth et al.\cite{NatCommun.7.13547} (Copyright © Nature Publishing Group).}
	\label{f8}
\end{figure*}

Similar with RMnO$_3$, INS data of this material also exhibits features originating from a magnon-phonon coupling as summarized in Fig. \ref{f7}:\cite{PhysRevB.94.104421} the downward shift at the M point and the relatively enhanced intensity of the top mode at around 12.5 meV. Other earlier INS studies tried to explain the downward shift using a large single-ion anisotropy.\cite{PhysRevB.84.094448,PhysRevB.81.104411} But this scenario, we think, is perhaps unrealistic, because the Cr$^{3+}$ ions at an octahedral site is supposed to have three $t_{2_g}$ quenched orbitals so that the exchange interaction is most likely  isotropic. Therefore, we propose that the magnon-phonon coupling Hamiltonian is necessary to fully account for the unconventional observed features. As explained in Sect. \ref{sec:2.2}, the coupling Hamiltonian based on the exchange-striction model was adapted for CuCrO$_2$ assuming one dispersionless optical phonon branch with an energy of 12.5 meV. It is also reasonable to assume that this phonon mode has stronger coupling with the observed magnons than other phonon modes, because the coupling strength is enhanced when the energies of quasiparticles are close with each other.\cite{RevModPhys.85.219}

As shown in Fig. \ref{f7}(c), using the coupling Hamiltonian we succeeded in reproducing the observed intensity of the 12.5 meV phonon mode. We further note that the observed intensity below T$_N$ at the M point was actually larger than the usual $q^2$ contribution from phonons. This intensity enhancement at the M point is also confirmed by a recent inelastic X-ray scattering (IXS) study.\cite{PhysRevB.95.054306} Such an intensity change occurs most probably through a spectral weight transfer originating from hybridization between magnons and phonons. The intensity calculated from this model is also consistent with the experimental observations, and shows stronger intensity around the M point. From this analysis, we found the coupling constant $c$ to be 16.8 meV/$\AA$, which can be converted into a dimensionless exchange-striction coefficient $\alpha$ = 15.8. Since the dominant interaction between Cr$^{3+}$ ions in this material is the direct exchange interaction,\cite{MRB.21.745} the value for CuCrO$_2$ is found to be largest among oxides studied so far (for the summary see Table \ref{t2}). 

\begin{table}[b!]
	\caption{Exchange-striction coefficients $\alpha$ for selected oxides}
	\begin{center}
		\begin{tabular}{ccc}
			\hline
			Materials&$\alpha$&References\\
			\hline
			CuGeO$_3$&3.5&Kodama et al.\cite{Science.298.395}\\
			La$_2$CuO$_4$&2$\sim$7&Hafliger et al.,\cite{PhysRevB.89.085113} Chernyshev et al.\cite{PhysRevB.92.054409}\\
			(Y,Lu)MnO$_3$&8$\sim$10&Oh et al.\cite{NatCommun.7.13146}\\
			HoMnO$_3$&12.8&Kim et al.\cite{PhysRevB.97.201113}\\
			CuCrO$_2$&15.8&Park et al.\cite{PhysRevB.94.104421}\\
			LiCrO$_2$&15.7&T\'oth et al.\cite{NatCommun.7.13547}\\
			\hline
		\end{tabular}
	\end{center}
	\label{t2}
\end{table}

\begin{figure*}
	\includegraphics[width=1\textwidth,clip]{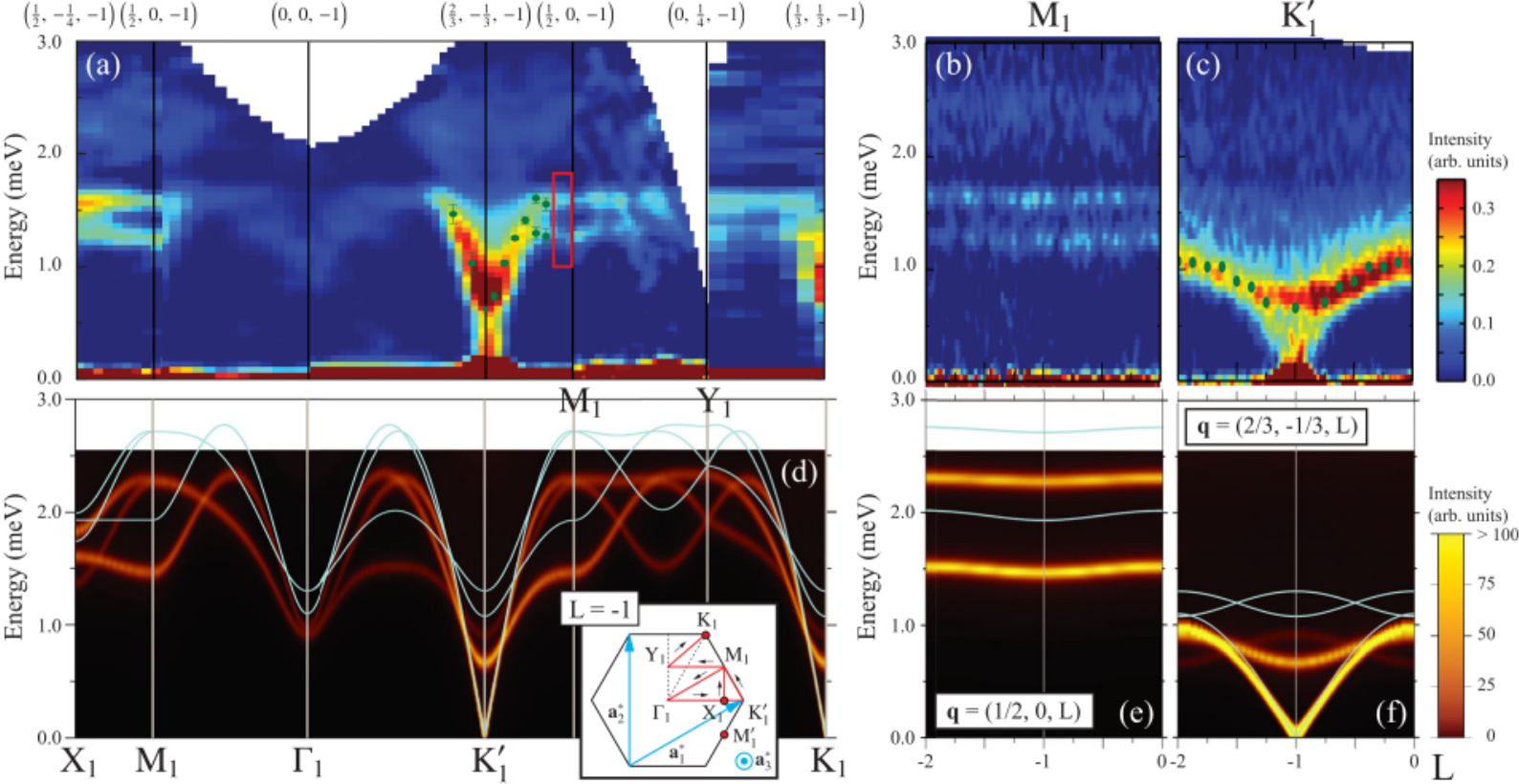}
	\caption{INS spectra of Ba$_3$CoSb$_2$O$_9$ as a function of the momentum and energy transfer at T = 1.5 K along the high symmetry (a) intralayer directions and the interlayer (b) $\left[1/2, 0, L \right]$, and (c) $\left[2/3, -1/3, L \right]$ directions in the reciprocal space. The red rectangular frame in (a) represents the region where the decay effect is distinct. Details are discussed in Ma et. al.\cite{PhysRevLett.116.087201} (d)(e)(f) The intensity plot of the dynamical structure factor along the same symmetry lines as in (a,b,c) for $J$ = 1.7 meV, $J_0/J$ = 0.05, and $\Delta$ = 0.89 at T = 0 calculated with the nonlinear spin wave approximation. The solid lines represent the poles in the LSWT approximation. Reprinted with permission from Ma et al.\cite{PhysRevLett.116.087201} (Copyright © American Physical Society).}
	\label{f9}
\end{figure*}

Using the magnon-magnon interaction, it is shown that the fitted two-magnon continuum signals, located between the 12.5 meV phonon and the 8 meV magnon, can be explained partly by the minimal spin Hamiltonian including higher order corrections. For example, the energy of observed and calculated two-magnon states are similar to one another, but the observed intensity is found to be slightly larger than the calculated one. Since the spin Hamiltonian alone cannot account for the contributions due to the decays from hybridized modes, we think that the calculated intensity is most likely to be underestimated. Thus, it is necessary to include the effects from coupling terms. Moreover, the nonlinear spin wave calculations for the lowest magnon mode do not show the roton-like minimum at the M point since the renormalization of the harmonic magnon energies are only 8 \% for the S=3/2 case.\cite{PhysRevB.88.094407} This clearly indicates that the magnon-magnon interaction alone cannot fully account for the observed minimum at the M point and therefore magnon-phonon coupling is necessary to explain the experimental observations. CuCrO$_2$ apparently has a stronger magnon-phonon coupling than magnon-magnon interaction.

As another example of delafossite systems, LiCrO$_2$ is also found to have a strong magnon-phonon coupling.\cite{NatCommun.7.13547} In LiCrO$_2$ case, the hybridized excitation was discovered by an IXS technique as shown in Fig. \ref{f8}(a). Upon cooling, a longitudinal acoustic phonon peaks loses its spectral weight, some of which is transferred to the new emergent peak located at around 10 meV. From the fully resolved dispersion of the hybridized excitations obtained from both IXS and INS, the roton-like minimum at the M point, the key signature of the magnon-phonon coupling, was also confirmed as plotted in Figs. \ref{f8}(b,c). It was noted as well that the hybridized excitation peak becomes very broad above T$_N$ (see Fig. \ref{f8}(a)). The authors suggested that the phonons ought to be coupled to the excitations of the magnetic correlated state, which persists above T$_N$ = 64 K due to the geomerical frustration in 2D TLAF.\cite{NatCommun.7.13547} The dispersion as well as the IXS cross section are found to be in good agreement with the model calculations accounting for the direct exchange-striction mechanism. Interestingly, the magnon-phonon coupling term accurately explains the minimum at the M$^\prime$ point and the strongest intensity at the $\Gamma$ and M$^\prime$ points, where the coupling term is estimated to be largest. The coupling strength determined from fitting is found to be 65 meV/$\AA$, which corresponds to an exchange-striction coefficient $\alpha$ = 15.7, a very similar value to that of CuCrO$_2$. We summarize the exchange-striction coefficients of some selected oxides in the Table \ref{t2}.

\subsection{2D TLAF with $S$ =  1/2 systems: Ba$_3$CoSb$_2$O$_9$ and Cs$_2$CuCl$_4$}
\label{sec:3.3}
So far, we examined the 2D TLAF with classical spin values where nonlinear effects are expected to be relatively weaker. In this regard, it is interesting to consider the $S$=1/2 2D TLAF because of their expected large quantum renormalization effect. Therefore, it is fortunate that the recently discovered compound Ba${}_{3}$CoSb${}_{2}$O${}_{9}$\cite{PhysRevLett.110.267201,PhysRevLett.109.267206,JPCM.16.8923} is a nearly ideal 2D TLAF system with spin $S$ =  1/2 and has a noncollinear magnetic structure.\cite{JPCM.16.8923} Here, the spin-orbit coupling induces a Kramers doublet ground state for Co${}^{2+}$ ions, resulting in an effective spin $S$ =  1/2 state. As theoretically studied,\cite{PhysRevB.79.144416,PhysRevLett.97.207202,RevModPhys.85.219,PhysRevB.88.094407} the magnon energy is markedly renormalized due to the enhanced magnon-magnon interactions. This was later experimentally confirmed by INS measurements as shown in Fig. \ref{f9}.\cite{PhysRevLett.116.087201} Adapting the Heisenberg $XXZ$ model including higher order corrections, the intra-plane and inter-plane nearest neighbor exchange interaction $J$ and $J'$ are estimated to be 1.7 and 0.085 meV, respectively. The estimate exchange anisotropy value of $\Delta=0.89$ makes it a rather anisotropic system. The observed renormalization value of the mode energies is found to be as much as about 40 \%, which is much larger than in most classical spin systems covered in this review such as LuMnO${}_{3}$ case (5 \%).\cite{PhysRevLett.111.257202} Thus, it is very clear that there are strong quantum renormalization effects in this compound.

A minimal $XXZ$ Hamiltonian including up to next-nearest neighbor exchange interaction was introduced to explain the experimental result by another group.\cite{PhysRevB.91.134423} However, the 1/S expansion of this Hamiltonian does not fully account for the following experimental findings: (1) along the $(00L)$ direction, the calculated intensity of magnon mode is overestimated, (2) the energy of upper magnon branch is overestimated, (3) the linewidth of magnon branch is broadened, even with a large anisotropy, $\Delta$ = 0.89. We note that below $\Delta$ = 0.92, the spontaneous magnon decay is not allowed theoretically. It implies that a new Hamiltonian term is needed to fully account for the observations. To explain these discrepancies, one may as well consider the magnon-phonon coupling for this quantum material. Indeed, the magnon-phonon coupling can renormalize the overestimated energy of the single magnon mode. However, since the energy scale of magnon in this material is quite small as compared with phonons, the coupling between magnons and phonons is thus expected to be relatively weak.

Despite such expectations, recent INS data obtained in the wide range of momentum-energy space revealed that there are multi-stage continuum states above the single magnon modes that have highly unusual dispersion curves\cite{NatCommun.8.235}, which were never reported before in other 2D TLAF. It was also noted that both the two-magnon continuum or two-spinon continuum\cite{PhysRevB.91.134423} are inadequate for explaining these multi-stage continua, because the energy range is extended up to six times above the range set by the exchange interaction. 

\begin{table*}[t!]
	\caption{{Summary table for the experimentally investigated TLAF materials}}
	\label{t3}
	\begin{center}
		\begin{tabular}{cccccc}
			\hline
			\multicolumn{1}{c}{Materials} & {$S$} &\multicolumn{1}{c}{Type of magnetic Hamiltonian} & \multicolumn{1}{c}{Features} & \multicolumn{1}{c}{Mechanism} & References \\
			\hline
			\multirow{2}*{h-LuFeO$_3$} & \multirow{2}*{5/2} & {Heisenberg XXZ } & \multirow{2}*{no anomalous features} & \multirow{2}*{-} & \multirow{2}*{Leiner et al.\cite{PhysRevB.98.134412}}\\
			&&with easy-plane anisotropy&&&\\
			\hline			
			\multirow{3}*{h-RMnO$_3$} & \multirow{3}*{2} & \multirow{2}*{Heisenberg XXZ} & {energy renormalization} & \multirow{2}*{exchange-striction} & {Oh et al.\cite{PhysRevLett.111.257202,NatCommun.7.13146}} \\
			& & & {hybridized mode and its decay} & & Kim et al.\cite{PhysRevB.97.201113}\\
			\cline{5-6}
			& & with easy-plane anisotropy & {downward shift at $M$ point} & {single-ion magneto-striction} & Holm et al.\cite{PhysRevB.97.134304}\\
			\hline			
			\multirow{2}*{ACrO$_2$} & \multirow{2}*{3/2} & \multirow{2}*{Heisenberg} & {hybridized mode} & {exchange-striction type} & T\"oth et al.\cite{NatCommun.7.13547} \\
			& & & {downward shift at $M$ point} & {magnon-phonon coupling} & Park et al.\cite{PhysRevB.94.104421}\\
			\hline
			\multirow{3}*{Ba$_3$CoSb$_2$O$_9$} & \multirow{3}*{1/2} & \multirow{3}*{Heisenberg XXZ} & {strong energy renormalization} & \multirow{2}*{magnon-magnon}& {Ma et al.\cite{PhysRevLett.116.087201}} \\
			& & &{intrinsic linewidth broadening} &  & {Ito et al.\cite{NatCommun.8.235}} \\ 
			& & & {large and broad continuum} & interaction & {Kamiya et al.\cite{NatCommun.9.2666}} \\
			\hline
			\multirow{3}*{Cs$_2$CuCl$_4$} & \multirow{3}*{1/2} & \multirow{3}*{Heisenberg} & {strong energy renormalization} & {magnon-magnon}& \multirow{2}*{Coldea et al.\cite{PhysRevLett.86.1335}} \\
			& & &{intrinsic linewidth broadening} & interaction &\\
			\cline{5-6}
			& & & {large and broad continuum} & spinon continuum  & Kohno et al.\cite{NatPhys.3.790}\\
			\hline				
		\end{tabular}
	\end{center}
\end{table*}

Other examples of 2D TLAF with $S$ =  1/2, include Cs${}_{2}$CuCl${}_{4}$ and Cs${}_{2}$CuBr${}_{4}$, which have a distorted 2D TLAF consisting of isosceles triangles due to the orthorhombic crystal structure. The space group of Cs${}_{2}$CuCl${}_{4}$ is \textit{Pnma} and the lattice constants are known to be $a$ = 9.65, $b$ = 7.48, and $c$ = 12.26 $\AA$ at 0.3 K.\cite{C.R.Acad.313.1149} Cu${}^{2+}$ ions represent spin $S$ =  1/2 and {CuCl$_4$}$^{2-}$ tetrahedrons form a triangular lattice within the $bc$ plane. The magnetic structure was found to have a cycloid state below T${}_{N}$ = 0.62 K at zero field.\cite{PhysRevLett.79.151,PhysRevB.68.134424,PhysRevLett.86.1335} The spin cycloids lie in the $bc$ plane with a propagation vector of $Q$~=~(0, 0.503, 0), with a small incommensurability due to the geometrical frustration. The spin Hamiltonian is defined by two main exchange interactions as follows: 
\[
H=J\sum^{\hat{b}}_{<i,i^{'}>}{\boldsymbol{S}_i\cdot\boldsymbol{S}_{i'}}+J'\sum^{\hat{c}}_{<i,j>}{\boldsymbol{S}_i\cdot\boldsymbol{S}_j}, \tag{25} \label{Cs2CuCl4}
\]
where $J$ and $J'$ denote the exchange interaction along the $b$ direction and the zig-zag bonds (along the $c$ axis), respectively. These coupling terms are not the same and found to be $J'/J$ = 0.33 with $J$ = 0.374 meV. The interlayer coupling $J''$ is not zero ($J''/J$ = 0.045), leading to the so-called cone state under external fields along the $a$ axis.\cite{PhysRevB.68.134424} Due to the small DM interaction ($D_a$/$J$ = 0.053), the spin cycloids are inclined away from the $bc$ plane. Cs${}_{2}$CuBr${}_{4}$, the isostructural compound of Cs${}_{2}$CuCl${}_{4}$, is found to have a rich phase diagram\cite{JPSJ.74.135,PhysRevLett.102.257201,PhysRevB.76.060406} and a different inter-chain coupling ($J'/J$~=~0.5).\cite{PhysRevB.67.104431,PhysRevB.71.134422} Below T${}_{N}$ = 1.4 K, the spins of Cs$_2$CuBr$_4$ are ordered like Cs${}_{2}$CuCl${}_{4}$ with a helical incommensurate structure $Q$~=~(0, 0.575, 0).

The magnetic excitations in Cs${}_{2}$CuCl${}_{4}$ obtained from INS have strong renormalization as compared to the LSWT prediction.\cite{PhysRevB.68.134424,PhysRevLett.86.1335} Interestingly, the large signals of the magnon continuum are observed to range up to a high energy transfer of $3J$. There are some features that cannot be fully explained by the two-magnon continuum: a long tail and a large spectral weight of continuum signals. The spin wave theory including $1/S$ expansions\cite{PhysRevB.73.184403,PhysRevB.72.134429} and the series expansion method\cite{PhysRevLett.96.057201,PhysRevB.75.174447} were used to account for those features. As shown in aforementioned noncollinear magnets, calculations for this case yield a roton-like minimum at the M point and a flat dispersion in the middle of the $\Gamma$K line.\cite{PhysRevB.79.144416,PhysRevB.74.180403,PhysRevB.74.224420} The decay of magnons from magnon-magnon interactions leads to a clear underlying picture of the renormalization and the strong broad continuum signals.

On the other hand, because of the anisotropic exchange interaction ($J'/J$ = 0.33), these systems can also be regarded as a 1D chain system with a weak inter-chain coupling. Therefore, the broad continuum state in the magnetic excitations observed in INS could be described as two-spinon continuum,\cite{NatPhys.3.790,PhysRevLett.96.057201} which is related to elementary fractionalized excitations of a typical 1D spin chain system. Indeed, the upper boundary of the continuum state and the calculated dynamical structure factor are found to be in good agreement with each other, while the low energy part is slightly different because the weak DM interaction was neglected in the calculation.

\section{Summary and Outlook}
\label{sec:4}
What has been shown in this review is that the fundamental quasiparticles of magnetic systems, magnons and phonons, do decay more often than not through either a nonlinear magnon coupling or a magnon-phonon coupling, sometimes called a spin-lattice coupling. The nonlinear magnon-magnon interaction is generally present for all magnetic systems. However, noncollinear magnetic systems provide the main feasible route for experimental measurements of the effects resulting from such interactions. At the same time, there are basically three mechanisms leading to magnon-phonon coupling: exchange-striction, antisymmetric exchange-striction, and single-ion magneto-striction (See the Table \ref{t3}, which summarizes the experimentally investigated TLAF systems). We have summarized the formalism of how one may calculate this magnon-phonon coupling. This is indeed a very general form that can be applied to any other magnetic materials. We note that both nonlinear magnon-magnon and magnon-phonon interactions provide the otherwise forbidden decay of single magnons. In a very loose analogy, this coupling between magnons and phonons is like the mixing of three types of neutrinos, so one can imagine having a certain magnon-phonon oscillation or conversion just as seen in neutrinos through the non-zero mixing angle.

We have laid out how to calculate the strength and properties of hybridized excitations for a given material with different degrees of magnon-phonon coupling constants. The understanding we have gained from this study may have implications in much wider fields. For example, understanding of how magnons decay through magnon-phonon and/or magnon-magnon couplings is closely related to some of the spintronic issues such as magnon damping or spin coherence time. More specifically, these couplings can also be found useful in understanding the spin Seebeck effect. 

One may also think of applying the concept of magnon-phonon coupling to the two very important fields like frustrated magnetism and colossal magnetoresistance (CMR), where the magnetic transition is often accompanied by or associated with a structural transition. On the more speculative side, one may also imagine that this renewed understanding of magnon-phonon coupling in noncollinear magnets can be useful for the study of the age-old Invar problem, in which certain materials exhibit either very small, zero, or sometimes negative thermal expansion. Because of the nature of this problem, there has been a long speculation that the Invar behavior may as well be somehow deeply connected to the coupling between spin and lattice degrees of freedom.\cite{DukJooKim} But to the best of our knowledge, there have been very few experimental or theoretical studies on how this idea may unravel the almost century old puzzle of the Invar problem. Looking ahead, it is going to be extremely challenging but important to extend the magnon-phonon coupling to two extreme cases of magnetism: the quantum spin case and the itinerant magnetism case.

\section{Acknowledgments}
We thank Joosung Oh for fruitful discussions and acknowledge Pyeongjae Park and Kyungsoo Kim for their useful comments. We would also like to acknowledge Sasha Chernyshev and Pavel Maksimov for their careful reading and comments. The work at the IBS (Institute for Basic Science) CCES (Center for Correlated Electron System) was supported by the research program of IBS (IBS-R009-G1).



\end{document}